\documentclass[preprint,aps,showpacs]{revtex4}
\usepackage{mathrsfs}
\usepackage{amsmath}
\usepackage{amssymb}
\usepackage{epsfig}
\usepackage{graphicx}
\usepackage{booktabs}
\usepackage{array}
\usepackage{multirow}
\usepackage{color}
\begin{document}
\renewcommand{\arraystretch}{0.5}
\newcommand{\beq}{\begin{eqnarray}}
\newcommand{\eeq}{\end{eqnarray}}
\newcommand{\non}{\nonumber\\ }

\newcommand{\acp}{ {\cal A}_{CP} }
\newcommand{\psl}{ p \hspace{-1.8truemm}/ }
\newcommand{\nsl}{ n \hspace{-2.2truemm}/ }
\newcommand{\vsl}{ v \hspace{-2.2truemm}/ }
\newcommand{\epsl}{\epsilon \hspace{-1.8truemm}/\,  }
\def \cpl{ Chin. Phys. Lett.  }
\def \ctp{ Commun. Theor. Phys.  }
\def \epjc{ Eur. Phys. J. C }
\def \jpg{  J. Phys. G }
\def \npb{  Nucl. Phys. B }
\def \plb{  Phys. Lett. B }
\def \prd{  Phys. Rev. D }
\def \prl{  Phys. Rev. Lett.  }
\def \zpc{  Z. Phys. C }
\def \jhep{ J. High Energy Phys.  }

\title{Improved Estimates of The $B_{(s)}\rightarrow V V$ Decays \\in Perturbative QCD Approach}
\author{Zhi-Tian Zou$^a$\footnote{zouzt@ytu.edu.cn }, Ahmed Ali$^b$\footnote{ahmed.ali@desy.de}, Cai-Dian L\"u $^c$\footnote{lucd@ihep.ac.cn}, Xin Liu$^d$, Ying Li$^a$\footnote{liying@ytu.edu.cn} }   \affiliation{a.~Department of Physics, Yantai University, Yantai 264005,China\\
b.~Theory Group, Deutsches Elektronen-Synchrotron DESY, D-22603 Hamburg, FRG\\
c.~Institute  of  High  Energy  Physics  and  Theoretical Physics Center for Science Facilities,
CAS, Beijing 100049, China\\
d.~School of Physics and Electronic Engineering, Jiangsu Normal University, Xuzhou 221116, China
}
\begin{abstract}
We reexamine the branching ratios, $CP$-asymmetries, and other observables in a large number of $B_q\to VV(q=u,d,s)$ decays in the perturbative QCD (PQCD) approach, where $V$ denotes a light vector meson $(\rho, K^*, \omega, \phi)$. The essential difference between this work and the earlier similar works is of parametric origin and in the estimates of the power corrections related to the ratio  $r_i^2=m_{V_i}^2/m_B^2(i=2,3)$ ($m_V$ and $m_B$ denote the masses of the vector and $B$ meson, respectively). In particular, we use up-to-date distribution amplitudes for the final state mesons and keep the terms proportional to the ratio $r_i^2$ in our calculations. Our updated calculations are in agreement with the experimental data, except for a limited number of decays which we discuss. We emphasize that the penguin annihilation and the hard-scattering emission contributions are essential to understand the polarization anomaly, such as in the $B\to \phi K^*$ and $B_s \to \phi\phi$ decay modes.  We also compare our results with those obtained in the QCD factorization (QCDF) approach and comment on the similarities and differences, which can be used to discriminate between these approaches in future experiments.
\end{abstract}
\preprint{DESY 15-001}
\maketitle
\section{Introduction}
Exclusive $B_q$ $(q=u,d,s)$ meson decays, especially  $B_q\to VV$ modes, where $V$ stands for a light vector  meson ($\rho, K^*, \omega, \phi$), have aroused a great deal of interest for both theorists~\cite{qcdf1,qcdf2,qcdf3,qcdfbtovv,qcdfbsvv,yang,btovv,bsvv,jpg32} and in experiments~\cite{hfag}. In contrast to the scalar and pseudoscalar mesons, vector mesons can be produced in several polarization states. Thus, the fraction of a given polarization state is an interesting observable, apart from the decay widths. Phenomenology of the $B_q\to VV$ decays offers rich opportunities for our understanding of the mechanism for hadronic weak decays and $CP$ asymmetry and searching for the effect of new physics beyond the standard model. In general, the underlying  dynamics for such decays is extremely complicated, but in the heavy quark limit  ($m_b\to \infty$), it is greatly simplified due to the factorization of the hadronic matrix elements in terms of the decay constants and form factors. Based on this, a number of two-body hadronic $B$ decays had been calculated in this so-called naive factorization approach~\cite{nfa}. However these calculations encounter three major difficulties: (i) for the so-called penguin-dominated, and also for the color-suppressed tree-dominated decays, the predicted branching ratios are systematically below the measurements, (ii) this approach can not account for the direct $CP$ asymmetries measured in experiments, and (iii) the predictions of transverse polarization fraction in penguin-dominated charmless $B_q\to VV$ decays are too small to explain the data, in which large such fractions are measured. All these indicate that this factorization approach needs improvements, for example by including some more perturbative QCD contributions~\cite{gaijin}. In the current market, there are essentially three approaches to implement perturbative improvements: QCD factorization (QCDF)~\cite{qcdf,pdm}, perturbative QCD approach(PQCD)~\cite{pqcd}, and the soft-collinear effective theory (SCET)~\cite{sect}. All these frameworks have been employed in the literature to quantitatively study the dynamics of the $B_q \to PP, VP, VV$ decays, having light pseudoscalar ($P$)and/or Vector ($V$) mesons in the final states.

In the $B_q\to VV$ decays, as the  $B_q$ meson is heavy, the vector mesons are energetic with $E_{V}\simeq m_{B}/2$. As the spectator quark ($u,d$ or $s$) in the $B_q$ meson is soft, a hard gluon exchange is needed to kick it into an energetic one to form a fast moving light vector meson. The theoretical picture here is that a hard gluon from the spectator quark connects with the other quarks of the four-quark operators of the weak interaction~\cite{pqcd}. The underlying theory is thus a six-quark effective theory, and can be perturbatively calculated~\cite{lv23275}. In contrast to the other two approaches (QCDF and SCET), the PQCD approach is based on the $k_T$ factorization formalism~\cite{pqcd1,pqcd2,pqcd3}. The basic idea here is to take into account the transverse momentum $k_T$ of the valence quarks in the hadrons, as a result of which the end-point singularity in the collinear factorization (employed in the QCDF approach) can be avoided. On the other hand, the transverse momentum dependence introduces an additional energy scale leading to double logarithms in QCD corrections. These terms could be resummed through the renormalization group approach, which results in the appearance of the Sudakov form factor. This form factor effectively suppresses the end-point contribution of the distribution amplitude of the mesons in the small transverse momentum region, making the calculation in the PQCD approach reliable. It is worth mentioning that in this framework, the so-called annihilation diagrams are also  perturbatively calculable without introducing additional parameters~\cite{anni1,anni2}. The PQCD approach has been successfully used to study a number of pure annihilation type decays, and these predictions were confirmed subsequently in experiments~\cite{bsvv,pipi,anni2,dk,anniexp}. Thus, in our view, this method is reliable in dealing with the pure annihilation-type and annihilation-dominated decays as well.

Several years ago, H. Y. Cheng and C. K. Chua updated~\cite{qcdfbtovv,qcdfbsvv} the previous predictions~\cite{qcdf1,qcdf2,qcdf3} for $B_q \to VV$ decays in the QCDF factorization approach by taking the transverse polarization contributions into account, and using the updated values of the parameters in the input wave functions and the form factors. In the PQCD framework, although many studies of the two-body $B_q$-decays are available~\cite{btovv,bsvv,jpg32}, a reappraisal is needed for the following reasons: (i) In the previous studies, the terms proportional to  ``$r_{i}^2=m_{V_i}^2/m_B^2\, (i=2,3)$" have been omitted in the amplitudes, especially in the denominator of the propagators of virtual quarks and gluons. As we point out later, these terms do bring the earlier PQCD predictions in better accord in terms of the measured observables in some problematic cases, such as the $B\to \phi K^*$ and $B_s \to \phi\phi$ decays, (ii) recent progress in the study of the distribution amplitudes of the vector meson, especially for the $\phi$ meson, undertaken in the context of the QCD sum rules, may significantly impact on some of the calculations done earlier, and (iii) Experimental data for some of the  $B_q\to VV$ decays, such as the branching ratio and the polarization fractions of $B_s \to \phi\phi$, are now available. In addition, we work out a number of observables, such as $\phi_{\parallel}$, $\phi_{\perp}$, $A_{CP}^0$, $A_{CP}^{\perp}$, $\Delta\phi_{\parallel}$ and $\Delta\phi_{\perp}$ for the first time in PQCD. Among others, we revisit the $B\to \rho~(\omega)\phi$ decay modes, the direct $CP$ asymmetry of which could help us distinguish the PQCD and competing approaches. A related issue is the large fraction of the transverse polarization observed in some of these  decays. In the PQCD framework, penguin-annihilation contribution is the key to understanding this phenomenon. Especially, the chirally enhanced (S-P)(S+P) penguin-annihilation gives rise to large transverse polarizations. Together with the hard spectator-scattering contributions, this could help solve the transverse polarization puzzle in the penguin-dominated $B_q \to VV$ decays.

This work is organized as follows. In Sec.~\ref{sec:function}, we outline the framework of the PQCD approach and specify the various input parameters, such as the wave functions and decay constants. Details of the perturbative calculations for the $B_q\to VV$ decays are presented in in Sec.~\ref{jiexi}, and the various input functions are given in the Appendix. Numerical results of our calculations are presented in Sec.~\ref{result}  and compared in detail with the available experiments and earlier theoretical works. Finally, a short summary is given in Sec.~\ref{summary}.

\section{FORMALISM AND WAVE FUNCTION}\label{sec:function}
Our goal is to calculate the transition matrix elements:
\begin{eqnarray}
\mathcal{M}\propto \langle VV|\mathcal{H}_{eff}|B_{q}\rangle~,
\end{eqnarray}
with the weak effective Hamiltonian $\mathcal{H}_{eff}$ written as \cite{rmp681125}
\begin{eqnarray}
\mathcal{H}_{eff}=\frac{G_{F}}{\sqrt{2}}\left\{ V_{ub}^{*}V_{uX}\left[C_{1}(\mu)O_{1}^{q}(\mu)+C_{2}(\mu)O_{2}^{q}(\mu)\right] -V_{tb}^{*}V_{tX}\left[\sum_{i=3}^{10}C_{i}(\mu)O_{i}(\mu)\right]\right\}~.
\label{eq:heff}
\end{eqnarray}
Here,  $V_{ub(X)}$ and $V_{tb(X)}$  ($X=d,s$) are the CKM matrix elements, $C_{i}(\mu)$ are the effective Wilson coefficient calculated at the scale $\mu$, and the local four-quark operators $O_{j}\,(j=1,...,10)$ are defined and classified as follows:
\begin{itemize}
\item Current-current (tree) operators,
\begin{eqnarray}
O_{1}^{u}=(\bar{b}_{\alpha}u_{\beta})_{V-A}(\bar{u}_{\beta}X_{\alpha})_{V-A},
\;\;\;O_{2}^{u}=(\bar{b}_{\alpha}u_{\alpha})
_{V-A}(\bar{u}_{\beta}X_{\beta})_{V-A},
\end{eqnarray}
\item QCD penguin operators,
\begin{eqnarray}
&&O_{3}=(\bar{b}_{\alpha}X_{\alpha})_{V-A}\sum_{q^{\prime}}(\bar{q}^{\prime}_{\beta}q^{\prime}_{\beta})_{V-A},\;\;\;
O_{4}=(\bar{b}_{\alpha}X_{\beta})_{V-A}\sum_{q^{\prime}}(\bar{q}^{\prime}_{\beta}q^{\prime}_{\alpha})_{V-A},\\
&&O_{5}=(\bar{b}_{\alpha}X_{\alpha})_{V-A}\sum_{q^{\prime}}(\bar{q}^{\prime}_{\beta}q^{\prime}_{\beta})_{V+A},\;\;\;
O_{6}=(\bar{b}_{\alpha}X_{\beta})_{V-A}\sum_{q^{\prime}}(\bar{q}^{\prime}_{\beta}q^{\prime}_{\alpha})_{V+A},
\end{eqnarray}
\item Electroweak penguin operators,
\begin{eqnarray}
&&O_{7}=\frac{3}{2}(\bar{b}_{\alpha}X_{\alpha})_{V-A}\sum_{q^{\prime}}e_{q^{\prime}}(\bar{q}^{\prime}_{\beta}q^{\prime}_{\beta})_{V+A},
\;\;O_{8}=\frac{3}{2}(\bar{b}_{\alpha}X_{\beta})_{V-A}\sum_{q^{\prime}}e_{q^{\prime}}(\bar{q}^{\prime}_{\beta}q^{\prime}_{\alpha})_{V+A},\\
&&O_{9}=\frac{3}{2}(\bar{b}_{\alpha}X_{\alpha})_{V-A}\sum_{q^{\prime}}e_{q^{\prime}}(\bar{q}^{\prime}_{\beta}q^{\prime}_{\beta})_{V-A},\;\;
O_{10}=\frac{3}{2}(\bar{b}_{\alpha}X_{\beta})_{V-A}\sum_{q^{\prime}}e_{q^{\prime}}(\bar{q}^{\prime}_{\beta}q^{\prime}_{\alpha})_{V-A},
\end{eqnarray}
\end{itemize}
with the SU(3) color indices $\alpha$ and $\beta$ and the active quarks $q^{\prime}=(u,d,s,c)$. The left-handed (right-handed) current $V\pm A$ are defined as $\gamma_{\mu}(1\pm\gamma_{5})$. Following~\cite{prd58094009}, we introduce the following combinations $a_{i}$ of the Wilson coefficients:
\begin{eqnarray}
&&a_{1}=C_{2}+C_{1}/3,\;\;\;\;\;\;a_{2}=C_{1}+C_{2}/3,\nonumber\\
&&a_{i}=C_{i}+C_{i\pm 1}/3,\,i=3,5,7,9\, / \,4,6,8,10.
\end{eqnarray}

In the perturbative approach to hadronic $B_q$ decays, several typical scales are encountered with large logarithms involving the ratios of these scales. They are resummed using the renormalization group (RG) techniques. Standard model specifies the Wilson coefficients at the electroweak scale $m_W$, the W boson mass, and the RG equations enable us to evaluate the dynamical effects in scaling the Wilson coefficients in Eq.~(\ref{eq:heff}) from $m_W$ to $m_b$, the $b$-quark mass. The physics between the scale $m_b$  and the factorization scale $\Lambda_h$, taken typically as $\Lambda_h \simeq \sqrt{m_b\Lambda_{\rm QCD}}$, can be calculated perturbatively and included in the so-called hard kernel in the PQCD approach. The soft dynamics below the factorization scale $\Lambda_h$ is nonperturbative and is described by the hadronic wave functions of the mesons involved in the decays $B_q \to VV$.  Finally, based on the factorization ansatz, the decay amplitudes are described by the convolution of the Wilson coefficients $C(t)$, the hard scattering kernel $H(x_i,b_i,t)$ and the light-cone wave functions $\Phi_{M_{i},B}(x_j,b_j)$ of the  mesons~\cite{prd55and56}:
\begin{eqnarray}
\mathcal
{A}\;\sim\;&&\int\,dx_{1}dx_{2}dx_{3}b_{1}db_{1}b_{2}db_{2}b_{3}db_{3}\nonumber\\
&&\times{\rm Tr}\left[C(t)\Phi_{B}(x_{1},b_{1})\Phi_{M_{2}}(x_{2},b_{2})\Phi_{M_{3}}
(x_{3},b_{3})H(x_{i},b_{i},t)S_{t}(x_{i})e^{-S(t)}\right],
\end{eqnarray}
where ${\rm Tr}$ denotes the trace over Dirac and color indices, $b_{i}$ are the conjugate variables of the quark transverse momenta $k_{iT}$, $x_{i}$ are the longitudinal momentum fractions carried by the quarks, and $t$ is the largest scale in the hard kernel $H(x_{i},b_{i},t)$. The jet function $S_{t}(x_{i})$ coming from the threshold resummation of the double logarithms $\ln^2 x_i$ smears the end-point singularities in $x_{i}$ \cite{prd66094010}. The Sudakov form factor  $e^{-S(t)}$ from the resummation of the double logarithms suppresses the soft dynamics effectively i.e. the long distance contributions in the large-$b$ region \cite{prd57443,lvepjc23275}.

In the PQCD approach, both the initial and the final state meson wave functions are important non-perturbative inputs. For $B_{q}~(q=u,d,s)$ meson, the light-cone matrix element could be decomposed as \cite{bwave1,bwave2}
\begin{eqnarray}
\int d^4z e^{ik\cdot z}\langle0|q_{\beta}(z)\bar{b}_{\alpha}(0)|B_{q}(P_{B_q})\rangle=
\frac{i}{\sqrt{6}}\left\{(\makebox[-1.5pt][l]{/}P_{B_q}+M_{B_q})\gamma_{5}\left[\phi_{B_q}(k)-
\frac{\makebox[-1.5pt][l]{/}n-\makebox[-1.2pt][l]{/}v}{\sqrt{2}\bar{\phi}_{B_q}}(k)\right]\right\}
_{\beta\alpha},
\end{eqnarray}
where $n=(1,0,\vec{0}_T)$ and $v=(0,1,\vec{0}_T)$ are the unit vectors of the light-cone coordinate system. Corresponding to the two Lorentz structures in the $B_{q}$ meson distribution amplitudes, there are two wave functions $\phi_{B_q}(k)$ and $\bar{\phi}_{B_q}(k)$, obeying the following normalization conditions:
\begin{eqnarray}
\int \frac{d^4k}{(2\pi)^4}\phi_{B_q}(k)=\frac{f_{B_q}}{2\sqrt{6}},\;\;\int\frac{d^4k}{(2\pi)^4}\bar{\phi}_{B_q}(k)=0,
\end{eqnarray}
where $f_{B_q}$ is the decay constant of the $B_q$ meson. Due to the numerical suppression, the contribution of $\bar{\phi}_{B}$  is often neglected. Finally, for convenience, the wave function of $B$ meson can be expressed as:
\begin{eqnarray}
\Phi_{B_{q}}(x,b)=\frac{i}{\sqrt{6}}(\makebox[-1.5pt][l]{/}P_{B_q}+M_{B_q})\gamma_{5} \phi_{B_q}(x,b),
\end{eqnarray}
with the light-cone distribution amplitude
\begin{eqnarray}
\phi_{B_q}(x,b)=N_{B_q}x^2(1-x^2)\exp\left[-\frac{M_{B_q}^2x^2}{2\omega_{q}}
-\frac{1}{2}w_{q}^2b^2\right],
\end{eqnarray}
where $N_{B_q}$ is a normalization factor and $\omega_{q}$ is a shape parameter. For $B^0$($B^{\pm}$) meson, we use $\omega_q=0.4\pm0.04$ GeV, which is determined by the calculation of form factor and other well known decay modes \cite{pqcd1,pqcd2,omg}. Taking into account the small SU(3) breaking and the fact that the $s$ quark is heavier than the $u$ or $d$ quark, we use the shape parameter $\omega_s=0.5\pm0.05$ GeV for the $B_s$ meson, indicating that the $s$ quark momentum fraction is larger than that of the $u$ or $d$ quark in the $B^{\pm}$ or $B^0$ meson \cite{bsvv}.

The light vector meson is treated as a light-light quark-antiquark system with the momentum $P^2=M_V^2$, and its polarization vectors $\epsilon$ include one longitudinal polarization vector $\epsilon_L$ and two transverse polarization vectors $\epsilon_T$, which are defined  in~\cite{btovv,bksphi}. Up to twist-3, the vector meson wave functions are given by \cite{vwave}:
\begin{eqnarray}
&&\Phi_{V}^{L}\,=\,\frac{1}{\sqrt{6}}\left[M_{V}\makebox[0pt][l]{/}
\epsilon_{L}\phi_{V}(x)\,+\,\makebox[0pt][l]{/}\epsilon_{L}\makebox[-1.5pt][l]
{/}P\phi_{V}^{t}(x)+M_{V}\phi_{V}^{s}(x)\right]\nonumber\\
&&\Phi_{V}^{\perp}\,=\,\frac{1}{\sqrt{6}}\left[M_{V}\makebox[0pt][l]{/}
\epsilon_{T}\phi_{V}^{v}(x)\,+\,\makebox[0pt][l]{/}\epsilon_{T}\makebox[-1.5pt][l]
{/}P\phi_{V}^{T}(x)\,+\,M_{V}i\epsilon_{\mu\nu\rho\sigma}\gamma_{5}
\gamma^{\mu}\epsilon_{T}^{\nu}n^{\rho}v^{\sigma}\phi_{V}^{a}(x)\right],
\end{eqnarray}
for the longitudinal polarization and the transverse polarization, respectively. Here $\epsilon_{\mu\nu\rho\sigma}$ is Levi-Civita tensor with the convention $\epsilon^{0123}=1$.

The twist-2 distribution amplitudes are given by
\begin{eqnarray}
&&\phi_{V}(x)\,=\,\frac{3f_{V}}{\sqrt{6}}x(1-x)\left[1+a^{\|}_{1V}C_{1}^{3/2}(t)
+a_{2V}^{\|}C_2^{3/2}(t)\right],\\
&&\phi_{V}^{T}(x)\,=\,\frac{3f_{V}^T}{\sqrt{6}}x(1-x)\left[1+a^{\perp}_{1V}C_{1}^{3/2}(t)
+a_{2V}^{\perp}C_2^{3/2}(t)\right],
\end{eqnarray}
with $t=2x-1$, and  $f_{V}^{(T)}$ are the decay constants of the vector meson, which for $V=\rho, \omega, K^*, \phi$ are shown numerically in Table \ref{tb:fv}. For the Gegenbauer moments, we use the following values\cite{vwave,jhep03}:
\begin{eqnarray}
&&a_{1\rho}^{\|(\perp)}=a_{1\omega}^{\|(\perp)}=a_{1\phi}^{\|(\perp)}=0,
\;\;a_{1K^*}^{\|(\perp)}=0.03\pm0.02\,(0.04\pm0.03)~,\nonumber\\
&&a_{2\rho}^{\|(\perp)}=a_{2\omega}^{\|(\perp)}=0.15\pm0.07\,(0.14\pm0.06)\;
a_{2\phi}^{\|(\perp)}=0\;(0.20\pm0.07)~,\nonumber\\
&&a_{2K^*}^{\|(\perp)}=0.11\pm0.09\;(0.10\pm0.08)~.
\end{eqnarray}
For the twist-3 distribution amplitudes, for simplicity, we adopt the asymptotic forms
\begin{eqnarray}
&&\phi_{V}^t(x)=\frac{3f_V^T}{2\sqrt{6}}t^2,\;\;\phi_{V}^s(x)=\frac{3f_{V}^T}{2\sqrt{6}}(-t),\nonumber\\
&&\phi_V^v(x)=\frac{3f_V}{8\sqrt{6}}(1+t^2),\;\;\phi_V^a(x)=\frac{3f_V}{4\sqrt{6}}(-t).
\end{eqnarray}

\begin{table}[htbp]
 \centering
 \caption{Input values of the decay constants of the light vector mesons, taken from~\cite{jhep03}}.
 \vspace{0.3cm}
 \begin{tabular}{c!{\;\;\;\;\;\;}c!{\;\;\;\;\;\;}c}
 \hline\hline
 \vspace{0.3cm}
  {vector} &  {$f_{V}$(MeV)} & {$f_{V}^{T}$(MeV)} \\
 \hline
 \vspace{0.1cm}
 {$\rho$}& {$216\,\pm\,3$}& {$165\,\pm\,9$}\\
\vspace{0.1cm}
 {$\omega$}& {$187\,\pm\,5$}& {$151\,\pm\,9$}\\
\vspace{0.1cm}
 {$K^*$}& {$220\,\pm\,5$}& {$185\,\pm\,10$}\\
\vspace{0.3cm}
 {$\phi$}& {$215\,\pm\,5$}& {$186\,\pm\,9$}\\
 \hline\hline
\end{tabular}\label{tb:fv}
\end{table}

\section{Perturbative calculation}\label{jiexi}

\begin{figure}[]
\begin{center}
\vspace{-2cm} \centerline{\epsfxsize=11 cm \epsffile{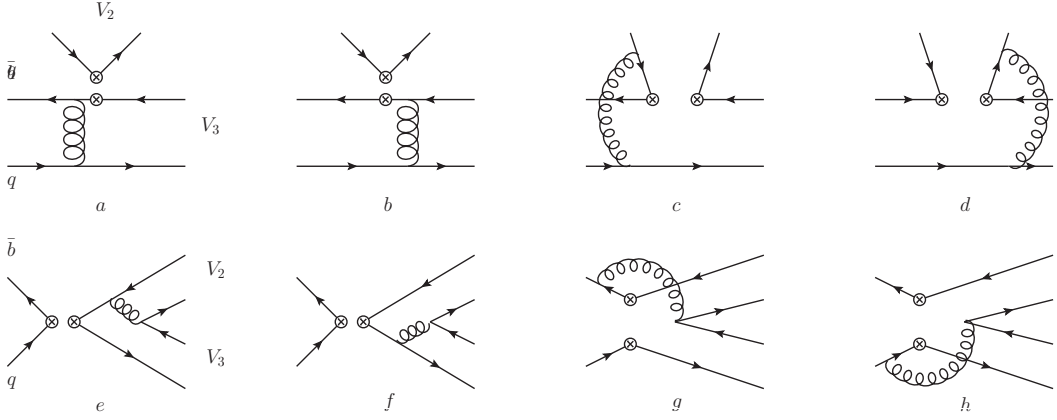}}
\vspace{-8cm} \caption{Leading order Feynman diagrams contributing to the
$B_{(s)}\,\rightarrow\,VV$ decays in PQCD }
 \label{fig:diagram}
 \end{center}
\end{figure}

At leading order, there are eight types of Feynman diagrams contributing to the $B_q\rightarrow VV$ decays, which are presented in Fig.\ref{fig:diagram}. The first row shows the emission-type diagrams, with the first two contributing to the usual form factor; the last two are the so-called hard-scattering emission diagrams. In fact, the first two diagrams are the only contributions calculated in the naive factorization approach. The second row shows the annihilation-type diagrams, with the first two factorizable and the last two nonfactorizable.

In the following, we shall give the general factorization amplitudes for these $B_q\rightarrow VV$ decays. We use the symbol $LL$ to describe  the amplitude of the $(V-A)(V-A)$ operators, $LR$ denotes the amplitude of the $(V-A)(V+A)$ operators and $SP$ denotes that of $(S-P)(S+P)$ operators resulting from the Fierz transformation of the $(V-A)(V+A)$ operators. For the $B_q\to VV$ decays, both the longitudinal polarization and the transverse polarization contribute. The amplitudes can be decomposed as follows:
\begin{eqnarray}
\mathcal {A}(\epsilon_{2},\epsilon_{3})=i\mathcal
{A}^{L}+i(\epsilon_{2}^{T*}\cdot\epsilon_{3}^{T*})\mathcal
{A}^{N}+(\epsilon_{\mu\nu\alpha\beta}n^{\mu}v^{\nu}\epsilon_{2}^{T*\alpha}\epsilon_{3}^{T*\beta})\mathcal
{A}^{T},
\label{amplitude}
\end{eqnarray}
where $\mathcal{A}^{L}$ is the longitudinally polarized decay amplitude, $\mathcal {A}^{T}$ and $\mathcal {A}^{N}$ are the transversely polarized contributions, and $\epsilon^T$ is the transverse polarization vector of the vector meson.

The longitudinal polarization amplitudes for the factorizable emission diagrams in Fig.(a) and (b) are as follows:
\begin{eqnarray}
\mathcal{A}_{ef}^{LL(LR),L}&=&-8\pi C_F M_{B}^4f_{V_2}\int_{0}^{1}dx_1dx_3\int_{0}^{1/\Lambda}b_1db_1b_3db_3\phi_{B}(x_1,b_1)\left\{\left[(-1+x_3)\phi_3(x_3)\right.\right.\nonumber\\
&&\left.\left.+r_3(2x_3-1)(\phi_3^s(x_3)+\phi_3^t(x_3))\right] E_{ef}(t_a)h_{ef}(x_1,x_3(1-r_2^2),b_1,b_3)\right.\nonumber\\
&&\left.+2r_3\phi_3^s(x_3) E_{ef}(t_b)h_{ef}(x_3,x_1(1-r_2^2),b_3,b_1)\right\},
\end{eqnarray}
where $r_{i}=\frac{M_{V_i}}{M_B}$ and $C_F=4/3$ is a color factor. The functions $h_{ef}$, $t_{a,b}$, and $E_{ef}$ can be found in Appendix \ref{app:a}. There is no $(S-P)(S+P)$ type amplitude, as a vector meson can not be produced through this type of operators. In the PQCD approach, the traditional emission contribution is also the dominant one. Unknown higher order perturbative QCD corrections will influence the emission contributions as well as those from other topologies.  At present, although the next-to-leading order (NLO) contributions have not been completed, the vertex correction has been done and is used to improve the predictions for the decays  $B \to \pi \rho(\omega)$  and the $B \to \pi$ form factors~\cite{zhourui}, which allows us to estimate the stability of the emission diagram in NLO. The results quoted below are based on the leading order calculations, but  we also estimate the uncertainties from the partial NLO contributions based on the available results, as explained in Sec.~\ref{result}.

The last two diagrams in the first row in Fig.\ref{fig:diagram} are the hard-scattering emission diagrams, whose contributions are given below:
\begin{eqnarray}
\mathcal{M}_{enf}^{LL,L}&=&-16\sqrt{\frac{2}{3}}\pi C_{F}M_{B}^{4}\int_{0}^{1}d[x]\int_{0}^{1/\Lambda}b_1db_1b_2db_2\phi_{B}(x_1,b_1)\phi_{2}(x_{2})\nonumber\\
&&\times \left\{\left[(x_2-1)\phi_3(x_3)+r_3x_3(\phi_3^s(x_3)-\phi_3^t(x_3))\right] E_{enf}(t_c)h_{enf}(\alpha,\beta_1,b_1,b_2)\right.\nonumber\\
&&\left.+\left[(x_2+x_3)\phi_3(x_3)-r_3x_3(\phi_3^s(x_3)+\phi_3^t(x_3))\right] E_{enf}(t_d)h_{enf}(\alpha,\beta_2,b_{1},b_{2})\right\},
\end{eqnarray}
\begin{eqnarray}
\mathcal{M}_{enf}^{LR,L}&=&16\sqrt{\frac{2}{3}}\pi C_{F}r_{2}M_{B}^{4}\int_{0}^{1}d[x]\int_{0}^{1/\Lambda}b_1db_1b_2db_2\phi_{B}(x_{1},b_1)\nonumber\\
&&\times\left\{\left[r_3((x_2-x_3-1)(\phi_2^s(x_2)\phi_3^s(x_3)-\phi_2^t(x_2)\phi_3^t(x_3))\right.\right.\nonumber\\
&&\left.\left.+(x_2+x_3-1)(\phi_2^t(x_2)\phi_3^s(x_3)-\phi_2^s(x_2)\phi_3^t(x_3)))\right.\right.\nonumber\\
&&\left.\left.+(x_2-1)\phi_{3}(x_3)(\phi_2^s(x_2)+\phi_2^t(x_2))\right] E_{enf}(t_c)h_{enf}(\alpha,\beta_1,b_1,b_2)\right.\nonumber\\
&&\left.+\left[r_3((x_3-x_2)(\phi_2^t(x_2)\phi_3^s(x_3)+\phi_2^s(x_2)\phi_3^t(x_3))\right.\right.\nonumber\\
&&\left.\left.+(x_2+x_3)(\phi_2^s(x_2)\phi_3^s(x_3)+\phi_2^t(x_2)\phi_3^t(x_3)))\right.\right.\nonumber\\
&&\left.\left.+x_2\phi_{3}(x_3)(\phi_2^s(x_2)-\phi_2^t(x_2))\right]E_{enf}(t_d)h_{enf}(\alpha,\beta_2,b_{1},b_{2})\right\},
\end{eqnarray}
\begin{eqnarray}
\mathcal{M}_{enf}^{SP,L}&=&-16\sqrt{\frac{2}{3}}\pi C_{F}M_{B}^{4}\int_{0}^{1}d[x]\int_{0}^{1/\Lambda}b_1db_1b_2db_2\phi_{B}(x_1,b_1)\phi_{2}(x_{2})\nonumber\\
&& \times \left\{\left[\phi_{3}(x_3)(x_2-x_3-1)+r_3x_3(\phi_{3}^s(x_3)+\phi_3^t(x_3))\right]E_{enf}(t_c)h_{enf}(\alpha,\beta_1,b_1,b_2)\right.\nonumber\\
&&\left.+\left[\phi_3(x_3)x_2+r_3x_3(\phi_3^t(x_3)-\phi_3^s(x_3))\right]E_{enf}(t_d)h_{enf}(\alpha,\beta_2,b_{1},b_{2})\right\}.
\end{eqnarray}
The functions $t_{c(d)}$, $E_{enf}$, $h_{enf}$, $\alpha$, $\beta_{i}$ for the nonfactorizable emission diagrams are also listed in Appendix \ref{app:a}. As is well known, the hard-scattering emission diagrams with a light meson (pseudoscalar or vector) are suppressed. This can be seen from the figures (c) and (d), which are symmetrical. But, compared with the figure (d), the anti-quark  propagator in figure (c) has an additional negative sign. As a result, the two contributions cancel each other.

Figures (e) and (f) are the factorizable annihilation diagrams, whose factorizable contributions are listed below:
\begin{eqnarray}
\mathcal{A}_{af}^{LL(LR),L}&=&8C_F\pi f_{B}M_{B}^{4}\int_{0}^{1}dx_2dx_3\int_{0}^{1/\Lambda}b_2db_2b_3db_3\nonumber\\
&&\times\left\{\left[\phi_{2}(x_2)\phi_3(x_3)(x_3-1)+2r_2r_3\phi_2^s(x_2)(x_3\phi_{3}^t(x_3)-(x_3-2)\phi_3^s(x_3))\right]\right.\nonumber\\
&&\left.\cdot E_{af}(t_e)h_{af}(\alpha_1,\beta,b_2,b_3)\right.\nonumber\\
&&-\left.\left[-x_2\phi_2(x_2)\phi_3(x_3)+2r_2r_3\phi_3^s(x_3)((x_2-1)\phi_2^t(x_2)+(x_2+1)\phi_2^s(x_2))\right]\right.\nonumber\\
&&\left.\cdot E_{af}(t_f)h_{af}(\alpha_2,\beta,b_3,b_2)\right\},
\label{eq:anlll}
\end{eqnarray}
\begin{eqnarray}
\mathcal{A}_{af}^{SP,L}&=&-16C_{F}f_{B}\pi M_{B}^{4}\int_{0}^{1}dx_2dx_3\int_{0}^{1/\Lambda}b_2db_2b_3db_3\nonumber\\
&&\times\left\{\left[2r_2\phi_3(x_3)\phi_{2}^{s}(x_2)+r_3(x_3-1)\phi_2(x_2)(\phi_{3}^s(x_3)+\phi_3^t(x_3))\right]\right.\nonumber\\
&&\left.\cdot E_{af}(t_e)h_{af}(\alpha_1,\beta,b_2,b_3)\right.\nonumber\\
&&\left.+\left[2r_3\phi_{2}(x_2)\phi_3^s(x_3)+r_2x_2\phi_3(x_3)(\phi_{2}^{t}(x_2)-\phi_2^s(x_2))\right]\right.\nonumber\\
&&\left.\cdot E_{af}(t_f)h_{af}(\alpha_2,\beta,b_3,b_2)\right\},
\label{eq:anspl}
\end{eqnarray}
and the related scales and the hard functions listed in Appendix~\ref{app:a}

The last two diagrams in Fig.1 are the nonfactorizable annihilation diagrams. The expressions for the corresponding amplitudes are as follows:
\begin{eqnarray}
\mathcal{M}_{anf}^{LL,L}&=&16\sqrt{\frac{2}{3}}C_F\pi M_{B}^{4}\int_{0}^{1}d[x]\int_{0}^{1/\Lambda}b_1db_1b_2db_2\phi_{B}(x_1,b_1)\nonumber\\
&&\times\left\{\left[r_2r_3(\phi_2^t(x_2)(\phi_3^t(x_3)(1-x_2+x_3)+\phi_3^s(x_3)(x_2+x_3-1))\right.\right.\nonumber\\
&&\left.\left.+\phi_2^s(x_2)(\phi_3^t(x_3)(1-x_2-x_3)+\phi_3^s(x_3)(x_2-x_3+3)))\right.\right.\nonumber\\
&&\left.\left.-x_2\phi_2(x_2)\phi_3(x_3)\right]E_{anf}(t_g)h_{anf}(\alpha,\beta_1,b_1,b_2)\right.\nonumber\\
&&-\left.\left[r_2r_3(\phi_{2}^{s}(x_2)(\phi_3^s(x_3)(1+x_2-x_3)+\phi_{3}^{t}(x_3)(x_2+x_3-1))\right.\right.\nonumber\\
&&\left.\left.+\phi_{2}^t(x_2)(\phi_3^s(x_3)(1-x_2-x_3)+\phi_3^t(x_3)(x_3-x_2-1)))\right.\right.\nonumber\\
&&\left.\left.+(x_3-1)\phi_2(x_2)\phi_3(x_3)\right]E_{anf}(t_h)h_{anf}(\alpha,\beta_2,b_1,b_2)\right\},
\end{eqnarray}
\begin{eqnarray}
\mathcal{M}_{anf}^{LR,L}&=&16\sqrt{\frac{2}{3}}C_F\pi M_{B}^{4}\int_{0}^{1}d[x]\int_{0}^{1/\Lambda}b_1db_1b_2db_2\phi_{B}(x_1,b_1)\nonumber\\
&&\times\left\{\left[\phi_3(x_3)(\phi_{2}^{t}(x_{2})+\phi_{2}^{s}(x_2))r_{2}(x_{2}-2)\right.\right.\nonumber\\
&&\left.\left.-\phi_{2}(x_{2})(\phi_3^s(x_3)-\phi_3^t(x_3))r_{3}(x_{3}+1)\right]E_{anf}(t_g)h_{anf}(\alpha,\beta_1,b_1,b_2)\right.\nonumber\\
&&\left.+\left[\phi_2(x_2)(\phi_{3}^{s}(x_{3})-\phi_{3}^{t}(x_3))r_{3}(x_{3}-1)\right.\right.\nonumber\\
&&\left.\left.-\phi_{3}(x_{3})(\phi_2^s(x_2)+\phi_2^t(x_2))r_{2}x_{3}\right]E_{anf}(t_h)h_{anf}(\alpha,\beta_2,b_1,b_2)\right\},
\end{eqnarray}
\begin{eqnarray}
\mathcal{M}_{anf}^{SP,L}&=&16\sqrt{\frac{2}{3}}C_F\pi M_{B}^{4}\int_{0}^{1}d[x]\int_{0}^{1/\Lambda}b_1db_1b_2db_2\phi_{B}(x_1,b_1)\nonumber\\
&&\times\left\{\left[r_2r_3(\phi_2^t(x_2)(\phi_3^t(x_3)(1-x_2+x_3)-\phi_3^s(x_3)(x_2+x_3-1))\right.\right.\nonumber\\
&&\left.\left.+\phi_2^s(x_2)(\phi_3^t(x_3)(x_2+x_3-1)+\phi_3^s(x_3)(x_2-x_3+3)))\right.\right.\nonumber\\
&&\left.\left.+(x_3-1)\phi_2(x_2)\phi_3(x_3)\right]E_{anf}(t_g)h_{anf}(\alpha,\beta_1,b_1,b_2)\right.\nonumber\\
&&-\left.\left[r_2r_3(\phi_{2}^{s}(x_2)(\phi_3^s(x_3)(1+x_2-x_3)+\phi_{3}^{t}(x_3)(1-x_2-x_3))\right.\right.\nonumber\\
&&\left.\left.+\phi_{2}^t(x_2)(\phi_3^s(x_3)(x_2+x_3-1)+\phi_3^t(x_3)(x_3-x_2-1)))\right.\right.\nonumber\\
&&\left.\left.-x_2\phi_2(x_2)\phi_3(x_3)\right]E_{anf}(t_h)h_{anf}(\alpha,\beta_2,b_1,b_2)\right\}.
\end{eqnarray}

For the $B_{(s)}\to VV$ decays, the transverse polarization amplitudes of the two factorizable emission diagrams yield:
\begin{eqnarray}
\mathcal{A}_{ef}^{LL(LR),N}&=&8\pi C_F M_{B}^4f_{V_2}r_2\int_{0}^{1}dx_1dx_3\int_{0}^{1/\Lambda}b_1db_1b_3db_3\phi_{B}(x_1,b_1)\left\{\left[\phi_3^T(x_3)\right.\right.\nonumber\\
&&\left.\left.+r_3((x_3+2)\phi_3^v(x_3)-x_3\phi_3^a(x_3))\right] E_{ef}(t_a)h_{ef}(x_1,x_3(1-r_2^2),b_1,b_3)\right.\nonumber\\
&&\left.+r_3(\phi_3^a(x_3)+\phi_3^v(x_3)) E_{ef}(t_b)h_{ef}(x_3,x_1(1-r_2^2),b_3,b_1)\right\},
\end{eqnarray}
\begin{eqnarray}
\mathcal{A}_{ef}^{LL(LR),T}&=&-8\pi C_F M_{B}^4f_{V_2}r_2\int_{0}^{1}dx_1dx_3\int_{0}^{1/\Lambda}b_1db_1b_3db_3\phi_{B}(x_1,b_1)\left\{\left[\phi_3^T(x_3)\right.\right.\nonumber\\
&&\left.\left.+r_3((x_3+2)\phi_3^a(x_3)-x_3\phi_3^v(x_3))\right] E_{ef}(t_a)h_{ef}(x_1,x_3(1-r_2^2),b_1,b_3)\right.\nonumber\\
&&\left.+r_3(\phi_3^a(x_3)+\phi_3^v(x_3)) E_{ef}(t_b)h_{ef}(x_3,x_1(1-r_2^2),b_3,b_1)\right\}.
\end{eqnarray}

The transverse polarization amplitudes of the two hard-scattering emission diagrams fig.(c) and (d) are given below:
\begin{eqnarray}
\mathcal{M}_{enf}^{LL,N}&=&16\sqrt{\frac{2}{3}}\pi C_{F}M_{B}^{4}r_2\int_{0}^{1}d[x]\int_{0}^{1/\Lambda}b_1db_1b_2db_2\phi_{B}(x_1,b_1)\nonumber\\
&&\times \left\{\left[(1-x_2)\phi_3^T(x_3)(\phi_2^a(x_2)+\phi_2^v(x_2))\right] E_{enf}(t_c)h_{enf}(\alpha,\beta_1,b_1,b_2)\right.\nonumber\\
&&\left.-\left[2r_3(x_2+x_3)(\phi_2^a(x_2)\phi_3^a(x_3)+\phi_2^v(x_2)\phi_3^v(x_3))\right.\right.\nonumber\\
&&\left.\left.-x_2\phi_3^T(x_3)(\phi_2^a(x_2)+\phi_2^v(x_2))\right] E_{enf}(t_d)h_{enf}(\alpha,\beta_2,b_{1},b_{2})\right\},
\end{eqnarray}
\begin{eqnarray}
\mathcal{M}_{enf}^{LL,T}&=&16\sqrt{\frac{2}{3}}\pi C_{F}M_{B}^{4}r_2\int_{0}^{1}d[x]\int_{0}^{1/\Lambda}b_1db_1b_2db_2\phi_{B}(x_1,b_1)\nonumber\\
&&\times \left\{\left[(x_2-1)\phi_3^T(x_3)(\phi_2^a(x_2)+\phi_2^v(x_2))\right] E_{enf}(t_c)h_{enf}(\alpha,\beta_1,b_1,b_2)\right.\nonumber\\
&&\left.+\left[2r_3(x_2+x_3)(\phi_2^a(x_2)\phi_3^v(x_3)+\phi_2^v(x_2)\phi_3^a(x_3))\right.\right.\nonumber\\
&&\left.\left.-x_2\phi_3^T(x_3)(\phi_2^a(x_2)+\phi_2^v(x_2))\right] E_{enf}(t_d)h_{enf}(\alpha,\beta_2,b_{1},b_{2})\right\},
\end{eqnarray}
\begin{eqnarray}
\mathcal{M}_{enf}^{LR,N}&=&16\sqrt{\frac{2}{3}}\pi C_{F}M_{B}^{4}\int_{0}^{1}d[x]\int_{0}^{1/\Lambda}b_1db_1b_2db_2\phi_{B}(x_1,b_1)\phi_2^T(x_2)\nonumber\\
&&\times \left\{\left[r_3x_3(\phi_3^a(x_3)-\phi_3^v(x_3))-\phi_3^T(x_3)(r_2^2(x_2-1)-x_3r_3^2)\right]\right.\nonumber\\
&&\left.\cdot E_{enf}(t_c)h_{enf}(\alpha,\beta_1,b_1,b_2)\right.\nonumber\\
&&\left.+\left[r_3x_3(\phi_3^a(x_3)-\phi_3^v(x_3))+\phi_3^T(x_3)(r_2^2x_2+r_3^2x_3))\right]\right.\nonumber\\
&&\left. E_{enf}(t_d)h_{enf}(\alpha,\beta_2,b_{1},b_{2})\right\},
\end{eqnarray}
\begin{eqnarray}
\mathcal{M}_{enf}^{LR,T}&=&16\sqrt{\frac{2}{3}}\pi C_{F}M_{B}^{4}\int_{0}^{1}d[x]\int_{0}^{1/\Lambda}b_1db_1b_2db_2\phi_{B}(x_1,b_1)\phi_2^T(x_2)\nonumber\\
&&\times \left\{\left[r_3x_3(\phi_3^v(x_3)-\phi_3^a(x_3))-\phi_3^T(x_3)(r_2^2(x_2-1)+x_3r_3^2)\right]\right.\nonumber\\
&&\left.\cdot E_{enf}(t_c)h_{enf}(\alpha,\beta_1,b_1,b_2)\right.\nonumber\\
&&\left.+\left[r_3x_3(\phi_3^v(x_3)-\phi_3^a(x_3))+\phi_3^T(x_3)(r_2^2x_2-r_3^2x_3))\right]\right.\nonumber\\
&&\left. E_{enf}(t_d)h_{enf}(\alpha,\beta_2,b_{1},b_{2})\right\},
\end{eqnarray}
\begin{eqnarray}
\mathcal{M}_{enf}^{SP,N}&=&16\sqrt{\frac{2}{3}}\pi C_{F}M_{B}^{4}r_2\int_{0}^{1}d[x]\int_{0}^{1/\Lambda}b_1db_1b_2db_2\phi_{B}(x_1,b_1)\nonumber\\
&&\times \left\{\left[2r_3(x_3+1-x_2)(\phi_2^v(x_2)\phi_3^v(x_3)-\phi_2^a(x_2)\phi_3^a(x_3))\right.\right.\nonumber\\
&&\left.\left.+(x_2-1)\phi_3^T(x_3)(\phi_2^v(x_2)-\phi_2^a(x_2))\right]E_{enf}(t_c)h_{enf}(\alpha,\beta_1,b_1,b_2)\right.\nonumber\\
&&\left.+\left[x_2\phi_3^T(x_3)(\phi_2^a(x_2)-\phi_2^v(x_2))\right]E_{enf}(t_d)h_{enf}(\alpha,\beta_2,b_{1},b_{2})\right\},
\end{eqnarray}
\begin{eqnarray}
\mathcal{M}_{enf}^{SP,T}&=&16\sqrt{\frac{2}{3}}\pi C_{F}M_{B}^{4}r_2\int_{0}^{1}d[x]\int_{0}^{1/\Lambda}b_1db_1b_2db_2\phi_{B}(x_1,b_1)\nonumber\\
&&\times \left\{\left[2r_3(x_2-x_3-1)(\phi_2^v(x_2)\phi_3^a(x_3)-\phi_2^a(x_2)\phi_3^v(x_3))\right.\right.\nonumber\\
&&\left.\left.+(x_2-1)\phi_3^T(x_3)(\phi_2^a(x_2)-\phi_2^v(x_2))\right]E_{enf}(t_c)h_{enf}(\alpha,\beta_1,b_1,b_2)\right.\nonumber\\
&&\left.+\left[x_2\phi_3^T(x_3)(\phi_2^v(x_2)-\phi_2^a(x_2))\right]E_{enf}(t_d)h_{enf}(\alpha,\beta_2,b_{1},b_{2})\right\}.
\end{eqnarray}

The transverse polarization amplitudes for the factorizable annihilation diagrams are:
\begin{eqnarray}
\mathcal{A}_{af}^{LL(LR),N}&=&8C_F\pi f_{B}r_2r_3M_{B}^{4}\int_{0}^{1}dx_2dx_3\int_{0}^{1/\Lambda}b_2db_2b_3db_3\nonumber\\
&&\times\left\{\left[(\phi_{2}^v(x_2)\phi_3^v(x_3)+\phi_2^a(x_2)\phi_3^a(x_3))(x_3-2)-\right.\right.\nonumber\\
&&\left.\left. x_3(\phi_2^v(x_2)\phi_{3}^a(x_3)+\phi_2^a(x_2)\phi_3^v(x_3))\right]E_{af}(t_e)h_{af}(\alpha_1,\beta,b_2,b_3)\right.\nonumber\\
&&+\left.\left[(x_2-1)(\phi_2^a(x_2)\phi_3^v(x_3)+\phi_2^v(x_2)\phi_3^a(x_3))+\right.\right.\nonumber\\
&&\left.\left.(x_2+1)(\phi_2^a(x_2)\phi_3^a(x_3)+\phi_2^v(x_2)\phi_3^v(x_2))\right]E_{af}(t_f)h_{af}(\alpha_2,\beta,b_3,b_2)\right\},
\label{eq:anlln}
\end{eqnarray}
\begin{eqnarray}
\mathcal{A}_{af}^{LR,T}&=&-\mathcal{A}_{af}^{LL,T}=8C_F\pi f_{B}r_2r_3M_{B}^{4}\int_{0}^{1}dx_2dx_3\int_{0}^{1/\Lambda}b_2db_2b_3db_3\nonumber\\
&&\times\left\{\left[(\phi_{2}^a(x_2)\phi_3^v(x_3)+\phi_2^v(x_2)\phi_3^a(x_3))(x_3-2)-\right.\right.\nonumber\\
&&\left.\left. x_3(\phi_2^v(x_2)\phi_{3}^v(x_3)+\phi_2^a(x_2)\phi_3^a(x_3))\right]E_{af}(t_e)h_{af}(\alpha_1,\beta,b_2,b_3)\right.\nonumber\\
&&+\left.\left[(x_2-1)(\phi_2^v(x_2)\phi_3^v(x_3)+\phi_2^a(x_2)\phi_3^a(x_3))+(x_2+1)\right.\right.\nonumber\\
&&\left.\left.\cdot (\phi_2^v(x_2)\phi_3^a(x_3)+\phi_2^a(x_2)\phi_3^v(x_2))\right]E_{af}(t_f)h_{af}(\alpha_2,\beta,b_3,b_2)\right\},
\label{eq:anllt}
\end{eqnarray}
\begin{eqnarray}
\mathcal{A}_{af}^{SP,N}&=&-\mathcal{A}_{af}^{SP,T}=16C_F\pi f_{B}M_{B}^{4}\int_{0}^{1}dx_2dx_3\int_{0}^{1/\Lambda}b_2db_2b_3db_3\nonumber\\
&&\times\left\{\left[r_2\phi_3^T(x_3)(\phi_{2}^a(x_2)+\phi_2^v(x_2))\right]E_{af}(t_e)h_{af}(\alpha_1,\beta,b_2,b_3)\right.\nonumber\\
&&-\left.\left[r_3\phi_2^T(x_2)(\phi_3^a(x_3)-\phi_3^v(x_3))\right]E_{af}(t_f)h_{af}(\alpha_2,\beta,b_3,b_2)\right\}.
\label{eq:anspt}
\end{eqnarray}
From Eqs.~(\ref{eq:anlll}) and (\ref{eq:anlln}), we find that large cancellations between the two annihilation type diagrams ( (e) and (f)) take place, as a result of which they are highly power suppressed. These two symmetric diagrams will cancel each other due to the relative negative sign introduced by the anti-quark propagator in diagram (e). This agrees with the naive argument that the annihilation contributions are negligible, especially for two identical final state mesons. For Eq.~(\ref{eq:anllt}), although the cancellations are not as severe as in Eq.~(\ref{eq:anlll}) and Eq.~(\ref{eq:anlln}), the contribution is also highly suppressed being proportional to $r_2r_3$. From Eqs.~(\ref{eq:anspl}) and (\ref{eq:anspt}), it is interesting to see that no cancellations or power suppression are involved. The chiral enhancement here is important to explain the large direct $CP$ asymmetry, generated by the strong phase and the transverse polarization fraction in the penguin-dominated $B$ decays~\cite{cptr}. The chirally enhanced penguin annihilation contribution will be discussed in Sec.\ref{result}.

For the nonfactorizable annihilation diagrams ((g) and (h)), we get:
\begin{eqnarray}
\mathcal{M}_{anf}^{LL,N}&=&\mathcal{M}_{anf}^{SP,N}=16\sqrt{\frac{2}{3}}C_F\pi M_{B}^{4}\int_{0}^{1}d[x]\int_{0}^{1/\Lambda}b_1db_1b_2db_2\phi_{B}(x_1,b_1)\nonumber\\
&&\times\left\{\left[-2r_2r_3(\phi_2^a(x_2)\phi_3^a(x_3)+\phi_2^v(x_2)\phi_3^v(x_3))\right.\right.\nonumber\\
&&\left.\left.-\phi_2^T(x_2)\phi_3^T(x_3)(r_2^2(x_2-1)-r_3^2x_3)\right]E_{anf}(t_g)h_{anf}(\alpha,\beta_1,b_1,b_2)\right.\nonumber\\
&&+\left.\left[\phi_2^T(x_2)\phi_3^T(x_3)(r_2^2x_2-r_3^2(x_3-1))\right]E_{anf}(t_h)h_{anf}(\alpha,\beta_2,b_1,b_2)\right\},
\end{eqnarray}
\begin{eqnarray}
\mathcal{M}_{anf}^{LL,T}&=&-\mathcal{M}_{anf}^{SP,T}=16\sqrt{\frac{2}{3}}C_F\pi M_{B}^{4}\int_{0}^{1}d[x]\int_{0}^{1/\Lambda}b_1db_1b_2db_2\phi_{B}(x_1,b_1)\nonumber\\
&&\times\left\{\left[2r_2r_3(\phi_2^a(x_2)\phi_3^v(x_3)+\phi_2^v(x_2)\phi_3^a(x_3))\right.\right.\nonumber\\
&&\left.\left.-\phi_2^T(x_2)\phi_3^T(x_3)(r_2^2(x_2-1)+r_3^2x_3)\right]E_{anf}(t_g)h_{anf}(\alpha,\beta_1,b_1,b_2)\right.\nonumber\\
&&+\left.\left[\phi_2^T(x_2)\phi_3^T(x_3)(r_2^2x_2+r_3^2(x_3-1))\right]E_{anf}(t_h)h_{anf}(\alpha,\beta_2,b_1,b_2)\right\},
\end{eqnarray}
\begin{eqnarray}
\mathcal{M}_{anf}^{LR,N}&=&-\mathcal{M}_{anf}^{LR,T}=16\sqrt{\frac{2}{3}}C_F\pi M_{B}^{4}\int_{0}^{1}d[x]\int_{0}^{1/\Lambda}b_1db_1b_2db_2\phi_{B}(x_1,b_1)\nonumber\\
&&\times\left\{\left[r_2(2-x_2)\phi_3^T(x_3)(\phi_2^a(x_2)+\phi_2^v(x_2))\right.\right.\nonumber\\
&&\left.\left.r_3(x_3+1)\phi_2^T(x_2)(\phi_3^a(x_2)-\phi_3^v(x_3))\right]
E_{anf}(t_g)h_{anf}(\alpha,\beta_1,b_1,b_2)\right.\nonumber\\
&&+\left.\left[r_3(x_3-1)\phi_2^T(x_2)(\phi_3^v(x_3)-\phi_3^a(x_3))+\right.\right.\nonumber\\
&&\left.\left.r_2x_2\phi_3^T(x_3)(\phi_2^a(x_2)+\phi_2^v(x_2))\right]
E_{anf}(t_h)h_{anf}(\alpha,\beta_2,b_1,b_2)\right\},
\end{eqnarray}
This completes the derivation of the various contributions in the $B_{(s)} \to VV$ decays. We now turn to the presentation of our numerical results in the next section.

\section{NUMERICAL RESULTS AND DISCUSSIONS}\label{result}
We start this section by listing the input parameters used in our numerical calculations. The vector meson decay constants have been summarized in Table \ref{tb:fv}. Other parameters, such as the CKM matrix elements, QCD scale (GeV), the masses (GeV) and the decay constant of the $B_{(s)}$ mesons (GeV) and the corresponding lifetimes (in $ps$) are taken from the PDG review~\cite{pdg} and are given below:
\begin{eqnarray}
&&\Lambda_{\overline{MS}}^{f=4}=0.25\pm0.05,\;M_{B}=5.279,\;M_{B_s}=5.366,\;f_{B}=0.21\pm0.02,\;f_{B_s}=0.24\pm0.03,\nonumber\\
&&\tau_{B^{\pm/0}}= 1.641/1.519,\; \tau_{B_s}= 1.497,\;m_{b}({\rm pole})=4.8,\nonumber\\
&&V_{ud}=0.97427\pm0.00015,\;V_{us}=0.22534\pm0.00065,\;V_{ub}=0.00351^{+0.00015}_{-0.00014},\nonumber\\
&&V_{td}=0.00867^{+0.00029}_{-0.00031},\;V_{ts}=0.0404_{-0.0005}^{+0.0011},\;V_{tb}=0.999146^{+0.000021}_{-0.000046},\nonumber\\
&&\alpha=(89_{-4.2}^{+4.4})^\circ,\;\gamma=(68_{-11}^{+10})^\circ.
\label{para}
\end{eqnarray}

With three polarization amplitudes, $\mathcal{A}^{L}$, $\mathcal{A}^{N}$, and $\mathcal{A}^{T}$, the decay width is expressed as
\begin{eqnarray}
\Gamma(B_{(s)}\rightarrow
VV)\,=\,\frac{|\overrightarrow{P}|}{8\pi
M_{B}^{2}}\left[\mid\mathcal{A}^{L}\mid^{2}+2(\mid\mathcal{A}^{N}\mid^{2}+\mid\mathcal{A}^{T}\mid^{2})\right],
\end{eqnarray}
where the analytic formulas for the amplitudes $\mathcal{A}^{L}$, $\mathcal{A}^{N}$ and $\mathcal{A}^{T}$ can be found in~\cite{bsvv,btovv,jpg32}. In our convention, given in  Eq.~(\ref{amplitude}), the helicity amplitudes are defined as follows:
\begin{eqnarray}
A_0=A_{L}=-\mathcal{A}^{L},\;\;\;A_{\parallel}=\sqrt{2}\mathcal{A}^{N},\;\;\;A_{\perp}=\sqrt{2}\mathcal{A}^{T},
\end{eqnarray}
where the definitions of $A_{0,\parallel,\perp}$ are the same as those in~\cite{bsvv}. In this work, we also predict their relative phases $\phi_{\parallel}=arg(A_{\parallel}/A_{0})$ and $\phi_{\perp}=arg(A_{\perp}/A_0)$. The polarization fractions $f_{L,\parallel,\perp}$ are defined as
\begin{eqnarray}
f_{L,\parallel,\perp}=\frac{\mid A_{L,\parallel,\perp}\mid^2}{\mid A_{0}\mid^{2}+\mid A_{\parallel}\mid^{2}+\mid A_{\perp}\mid^{2}}~.
\end{eqnarray}
In addition to the direct $CP$ asymmetry parameters, we also evaluate the following observables:
\begin{eqnarray}
 A_{CP}^0=(\bar{f}_L-f_L)/(\bar{f}_L+f_L),\;\;\;
 A_{CP}^{\perp}=(\bar{f}_{\perp}-f_{\perp})/(\bar{f}_{\perp}+f_{\perp}),\;\;\;
 \Delta\phi_{\parallel}=(\bar{\phi}_{\parallel}-\phi_{\parallel})/2~.
\end{eqnarray}
 Of these, $\Delta\phi_{\perp}=(\bar{\phi}_{\perp}-\phi_{\perp})/2$ is being worked out for the $B_{q}\to VV$ decays for the first time.

\begin{table}[!t]
\centering
 \caption{Updated branching ratios, percentage of the longitudinal polarization $f_{L}$ and
the  transverse polarizations $f_{\perp}$, relative phases, and the $CP$ asymmetry parameters $A^{0}_{CP}$ and $A^{\perp}_{CP}$ in the
$B \rightarrow K^{*0}\phi$,
 $B_s \rightarrow \bar{K}^{*0}\phi$, $B_s\to \phi\phi$ and $B_s\to \bar{K}^{*0}K^{*0}$ decays calculated in the PQCD approach.}
 \vspace{0.cm}
\begin{tabular}[t]{lcccccccccc}
\hline\hline
 Modes& $Br(10^{-6})$ &$f_{L}$(\%) &$f_{\perp}$ (\%)& $\phi_{\parallel}$(rad)& $\phi_{\perp}$(rad)\\
 \hline
{$B^0\rightarrow K^{*0}\phi$}
&{$9.8_{-3.8}^{+4.9}$}
&{$56.5_{-5.9}^{+5.8}$}
&{$21.3_{-2.9}^{+2.8}$}
&{$2.15_{-0.19}^{+0.22}$}
&{$2.14_{-0.19}^{+0.23}$}\\
{$Exp$}
&{$9.8\pm0.6$}
&{$48\pm3$}
&{$24\pm5$}
&{$2.40\pm0.13$}
&{$2.39\pm0.13$}\\
\hline
 {$B^+\to K^{*+}\phi$}&  {$10.3_{-3.8}^{+4.9}$} &  {$57.0_{-5.9}^{+6.3}$} & {$21.0_{-3.0}^{+3.0}$}& {$2.18_{-0.19}^{+0.23}$}& {$2.19_{-0.20}^{+0.22}$}\\
\vspace{0.13cm}
 {$Exp$}&  {$10.0\pm2.0$} &  {$50\pm5$} & {$20\pm5$}& {$2.34\pm0.18$}& {$2.58\pm0.17$}\\
\vspace{0.13cm}
 {$B_s\to \phi\phi$}  &  {$16.7_{-7.1}^{+8.9}$} &  {$34.7_{-7.1}^{+8.9}$} & {$31.6_{-4.4}^{+3.5}$}& {$2.01_{-0.23}^{+0.23}$}& {$2.00_{-0.21}^{+0.24}$}\\
\vspace{0.13cm}
 {$Exp$}  &  {$19\pm5$} &  {$34.8\pm4.6$} &  {$36.5\pm4.4\pm2.7$ }& {$2.71^{+0.31}_{-0.36}\pm0.22$}& {$$}\\
\vspace{0.13cm}
 {$B_s\to \bar{K}^{*0}\phi$}  &  {$0.39_{-0.17}^{+0.20}$} &  {$50.0_{-7.2}^{+8.1}$}& {$24.2_{-3.9}^{+3.6}$}& {$1.95_{-0.22}^{-0.21}$}& {$1.95_{-0.22}^{+0.21}$}\\
\vspace{0.13cm}
 {$Exp$\footnotemark[1]}  &  {$1.10\pm0.29$} &  {$51\pm15\pm7$}& {$28\pm11\pm2$}& {$1.75\pm0.58\pm0.30$}& {$$}\\
\vspace{0.13cm}
 {$B_s\to K^{*0}\bar{K}^{*0}$} &  {$5.4_{-2.4}^{+3.0}$} &  {$38.3_{-10.5}^{+12.1}$} & {$30.0_{-6.1}^{+5.3}$} & {$2.12_{-0.25}^{+0.21}$}& {$2.15_{-0.23}^{+0.22}$}\\
\vspace{0.3cm}
 {$Exp$} &  {$28.1\pm4.6\pm5.6$} &  {$31\pm12\pm4$} &  {$38\pm11\pm4$} & {$$}& {$$}\\
\hline
&$A^{dir}_{CP}$(\%)&$A^{0}_{CP}$(\%)&$A^{\perp}_{CP}$(\%)&$\Delta\phi_{\parallel}(rad)$&$\Delta\phi_{\perp}(rad)$&\\
\hline
\vspace{0.13cm}
 {$B^0\to K^{*0}\phi$}& {$0.0$} & {$0.0$} & {$0.0$}& {$0.0$}& {$0.0$}\\
\vspace{0.13cm}
 {$Exp$}& {$$} & {$4\pm6$} & {$-11\pm12$}& {$0.11\pm0.22$}& {$0.08\pm0.22$}\\
\vspace{0.13cm}
 {$B^+\to K^{*+}\phi$}&  {$-1.0_{-0.26}^{+0.18}$} &  {$-0.60_{-0.14}^{+0.12}$} & {$0.75_{-0.11}^{+0.23}$}& {$-0.05_{-0.33}^{+0.12}$}& {$-0.01$}\\
\vspace{0.13cm}
 {$Exp$}&  {$-1\pm8$} &  {$17\pm11\pm2$} & {$22\pm24\pm8$}& {$0.07\pm0.2\pm0.05$}& {$0.19\pm0.20\pm0.07$}\\
\vspace{0.13cm}
 {$B_s\to \phi\phi$}  &  {$0.0$} &  {$0.0$} & {$0.0$}& {$0.0$}& {$0.0$}\\
\vspace{0.13cm}
 {$B_s\to \bar{K}^{*0}\phi$}  &  {$0.0$} &  {$0.0$}& {$0.0$}& {$0.0$}& {$0.0$}\\
\vspace{0.3cm}
 {$B_s\to K^{*0}\bar{K}^{*0}$} &  {$0.0$} &  {$0.0$} &  {$0.0$} & {$0.0$}& {$0.0$}\\
 \hline\hline
\end{tabular}\label{tb:zui}
\footnotetext[1]{The experimental results are taken from~\cite{bsksphi}.}
\end{table}

\begin{table}[!t]
\centering
 \caption{Updated branching ratios (in units of $10^{-6}$) of $B\rightarrow VV$ decays calculated in the PQCD approach. For comparison, we also give the updated theoretical predictions in the QCD factorization (QCDF) approach \cite{qcdfbtovv} and the previous predictions in the PQCD approach\cite{btovv}. Experimental data are from the Particle Data Group~\cite{pdg}}
 \vspace{0.1cm}
\begin{tabular}[t]{lccccc}
\hline\hline
\vspace{0.3cm}
 {Decay Modes}&  {Class} &  {This work}& { QCDF}& { PQCD(former)}&  {Exp}\\
 \hline
\vspace{0.2cm}
 {$B^0\rightarrow \rho^{0}\rho^{0}$}& {C} & {$0.27_{-0.09-0.04-0.01}^{+0.10+0.06+0.00}$}& {$0.9_{-0.4-0.2}^{+1.5+1.1}$}& {$0.9\pm0.1\pm0.1$}& {$0.73\pm0.28$} \\
\vspace{0.2cm}
 {$B^{0}\rightarrow \rho^{+}\rho^{-}$}&  {T}& {$26.0_{-8.1-1.4-1.2}^{+10.1+1.4+1.5}$}& {$25.5_{-2.6-1.5}^{+1.5+1.1}$}& {$35\pm5\pm4$}& {$24.2\pm3.1$} \\
\vspace{0.2cm}
 {$B^{0}\rightarrow \rho^{0}\omega$} &  {E,P} & {$0.40_{-0.12-0.08-0.01}^{+0.15+0.09+0.01}$} & {$0.08_{-0.02-0.00}^{+0.02+0.36}$}& {$1.9\pm0.2\pm0.2$}& {$<1.6$} \\
\vspace{0.2cm}
 {$B^{0}\rightarrow \omega\omega$} &  {C,P} & {$0.50_{-0.18-0.07-0.05}^{+0.21+0.09+0.05}$} & {$0.7_{-0.3-0.2}^{+0.9+0.7}$}& {$1.2\pm0.2\pm0.2$}& {$<4.0$} \\
\vspace{0.2cm}
 {$B^{0}\rightarrow K^{*0}\rho^{0}$} &  {P} &  {$3.3_{-1.1-0.9-0.1}^{+1.3+1.1+0.0}$} & {$4.6_{-0.5-3.5}^{+0.6+3.5}$}& {$5.9$}& {$3.4^{+1.7}_{-1.3}$} \\
\vspace{0.2cm}
 {$B^{0}\rightarrow K^{*+}\rho^{-}$} &  {P} &  {$8.4_{-2.8-1.9-0.9}^{+3.1+2.2+0.6}$} & {$8.9_{-1.0-5.5}^{+1.1+4.8}$}& {13}& {$<12.0$} \\
\vspace{0.2cm}
 {$B^0\rightarrow K^{*0}\omega$} &  {P} &  {$4.7_{-1.5-1.3-0.3}^{+2.1+1.6+0.2}$} & {$2.5_{-0.4-1.5}^{+0.4+2.5}$}& {9.6}& {$2.0\pm0.5$} \\
\vspace{0.2cm}
 {$B^0\rightarrow K^{*0}\bar{K}^{*0}$} &  {P} & {$0.34_{-0.11-0.09-0.03}^{+0.13+0.10+0.02}$} & {$0.6_{-0.1-0.3}^{+0.1+0.2}$}& {0.35}& {$0.8\pm0.5$} \\
\vspace{0.2cm}
 {$B^0\rightarrow K^{*+}K^{*-}$} &  {E} & {$0.21_{-0.09-0.05-0.02}^{+0.09+0.03+0.01}$} & {$0.1_{-0.0-0.1}^{+0.0+0.1}$}& {0.11}& {$<2.0$} \\
\vspace{0.2cm}
 {$B^0\rightarrow \rho^0\phi$} &  {P} & {$0.013_{-0.006-0.002-0.001}^{+0.007+0.001+0.001}$} &&& {$<0.33$} \\
\vspace{0.2cm}
 {$B^0\rightarrow \omega\phi$} &  {P} & {$0.010_{-0.004-0.002-0.001}^{+0.005+0.001+0.001}$} &&& {$<1.2$} \\
\vspace{0.2cm}
 {$B^0\rightarrow \phi\phi$} &  {P} & {$0.012_{-0.002-0.004-0.001}^{+0.003+0.005+0.001}$} && {$0.0189_{-0.0021}^{+0.0061}$}& {$<0.2$} \\
\vspace{0.2cm}
 {$B^+\rightarrow \rho^{+}\rho^{0}$} &  {T} &  {$13.5_{-3.9-0.7-1.0}^{+5.0+0.4+1.1}$} & {$20.0_{-1.9-0.9}^{+4.0+2.0}$}& {$17\pm2\pm1$}& {$24.0\pm1.9$} \\
\vspace{0.2cm}
 {$B^+\rightarrow \rho^{+}\omega$} &  {T} &  {$12.1_{-3.7-0.4-0.1}^{+4.5+0.1+0.1}$} & {$16.9_{-1.6-0.9}^{+3.2+1.7}$}& {$17\pm2\pm1$}& {$15.9\pm2.1$} \\
\vspace{0.2cm}
 {$B^+\rightarrow \rho^{+}K^{*0}$} &  {P} &  {$9.9_{-3.3-2.4-0.5}^{+3.8+2.7+0.3}$} & {$9.2_{-1.1-5.4}^{+1.2+3.6}$}& {$17$}& {$9.2\pm1.5$} \\
\vspace{0.2cm}
 {$B^+\rightarrow \rho^{0}K^{*+}$} &  {P} &  {$6.1_{-1.9-1.3-0.5}^{+2.5+1.3+0.3}$} & {$5.5_{-0.5-2.5}^{+0.6+1.3}$}& {$9.0$}& {$4.6\pm1.1$} \\
\vspace{0.2cm}
 {$B^+\rightarrow \omega K^{*+}$} &  {P} &  {$4.0_{-1.3-0.9-0.3}^{+1.7+1.3+0.3}$} & {$3.0_{-0.3-1.5}^{+0.4+2.5}$}& {$7.9$}& {$<7.4$} \\
\vspace{0.2cm}
 {$B^+\rightarrow K^{*+}\bar{K}^{*0}$} &  {P} & {$0.56_{-0.19-0.12-0.02}^{+0.23+0.13+0.02}$} & {$0.6_{-0.1-0.3}^{+0.1+0.3}$}& {$0.40$}& {$1.2\pm0.5$} \\
\vspace{0.4cm}
 {$B^{+}\rightarrow \rho^{+}\phi$} &  {P} & {$0.028_{-0.012-0.004-0.002}^{+0.015+0.003+0.002}$} & && {$<3.0$}\\
 \hline\hline
\end{tabular}\label{tb:br}
\end{table}

\begin{table}[!t]
\centering
 \caption{Updated percentage of the longitudinal polarizations $f_{L}$ of $B\rightarrow VV$ decays calculated
 in the PQCD approach compared with the updated theoretical predictions in the QCD factorization (QCDF) approach \cite{qcdfbtovv} and the previous predictions in the PQCD approach~\cite{btovv}. Experimental data are from the Particle Data Group\cite{pdg}. }
 \vspace{0.2cm}
\begin{tabular}[t]{lcccc}
\hline\hline
\vspace{0.25cm}
  {Decay Modes} &  { This work }&  {QCDF}&  {PQCD(former)}& {Exp.}\\
 \hline
\vspace{0.13cm}
 {$B^0\rightarrow \rho^{0}\rho^{0}$}& {$0.12_{-0.02-0.01-0.00}^{+0.04+0.15+0.00}$} & {$0.92_{-0.04-0.37}^{+0.03+0.06}$} & {$0.60$}& {$0.75\pm0.14$\footnotemark[1]}\\
\vspace{0.13cm}
 {$B^0\rightarrow \rho^{+}\rho^{-}$}&  {$0.95_{-0.01-0.01-0.00}^{+0.01+0.01+0.00}$} & {$0.92_{-0.02-0.02}^{+0.01+0.01}$} & {$0.94$}& {$0.977\pm0.026$}\\
\vspace{0.13cm}
 {$B^0\rightarrow \rho^{0}\omega$}  &  {$0.67_{-0.06-0.04-0.06}^{+0.04+0.03+0.06}$} & {$0.52_{-0.25-0.36}^{-0.11+0.50}$} &  {$0.87$}&\\
\vspace{0.13cm}
 {$B^0\rightarrow \omega\omega$}  &  {$0.66_{-0.10-0.02-0.04}^{+0.07+0.04+0.06}$} & {$0.94_{-0.01-0.20}^{+0.01+0.04}$}&  {$0.82$}&\\
\vspace{0.13cm}
 {$B^0\rightarrow K^{*0}\rho^{0}$} &  {$0.65_{-0.03-0.04-0.00}^{+0.03+0.03+0.00}$} & {$0.39_{-0.00-0.31}^{+0.00+0.60}$} &  {$0.74$} &  {$0.57\pm0.10$}\\
\vspace{0.13cm}
 {$B^0\rightarrow K^{*+}\rho^{-}$} &  {$0.68_{-0.03-0.03-0.02}^{+0.04+0.03+0.02}$} & {$0.53_{-0.03-0.32}^{+0.02+0.45}$} &  {$0.78$} &  {$$}\\
\vspace{0.13cm}
 {$B^0\rightarrow K^{*0}\omega$} &  {$0.65_{-0.05-0.02-0.00}^{+0.05+0.02+0.00}$} & {$0.58_{-0.10-0.14}^{+0.07+0.43}$} &  {$0.82$} &  {$0.69\pm0.13$}\\
\vspace{0.13cm}
 {$B^0\rightarrow K^{*0}\bar{K}^{*0}$} &  {$0.58_{-0.08-0.02-0.01}^{+0.07+0.02+0.02}$} & {$0.52_{-0.07-0.48}^{+0.04+0.48}$} &  {$0.78$} &  {$0.80\pm0.13$}\\
\vspace{0.13cm}
 {$B^0\rightarrow K^{*+}K^{*-}$} &  {$\sim1.0$} &  {$\sim1.0$} &  {$0.99$} &  {$$}\\
\vspace{0.13cm}
 {$B^0\rightarrow \rho^{0}\phi$} &  {$0.95_{-0.01-0.01-0.00}^{+0.01+0.01+0.00}$} & {$$} & {$$} & {$$}\\
\vspace{0.13cm}
 {$B^0\rightarrow \omega\phi$} &  {$0.94_{-0.02-0.02-0.00}^{+0.02+0.01+0.00}$} &  {$$} & {$$} & {$$}\\
\vspace{0.13cm}
 {$B^0\rightarrow \phi\phi$} &  {$0.97_{-0.01-0.01-0.00}^{+0.01+0.01+0.00}$} &  {$$} & {$0.65$} & {$$}\\
\vspace{0.13cm}
 {$B^+\rightarrow \rho^{+}\rho^{0}$} &  {$0.98_{-0.01-0.01-0.00}^{+0.01+0.01+0.00}$} & {$0.96_{-0.01-0.02}^{+0.01+0.02}$} &  {$0.94$} &  {$0.95\pm0.016$}\\
\vspace{0.13cm}
 {$B^+\rightarrow \rho^{+}\omega$} &  {$0.97_{-0.01-0.00-0.00}^{+0.01+0.00+0.00}$} & {$0.96_{-0.01-0.03}^{+0.01+0.02}$} &  {$0.97$} &  {$0.90\pm0.06$}\\
\vspace{0.13cm}
 {$B^+\rightarrow K^{*+}\rho^0$} &  {$0.75_{-0.03-0.03-0.02}^{+0.03+0.02+0.02}$} & {$0.67_{-0.03-0.48}^{+0.02+0.31}$} &  {$0.85$} &  {$0.78\pm0.12$}\\
\vspace{0.13cm}
 {$B^+\rightarrow K^{*0}\rho^+$} &  {$0.70_{-0.03-0.04-0.01}^{+0.03+0.04+0.00}$} & {$0.48_{-0.04-0.40}^{+0.03+0.52}$\;\footnotemark[2]} &  {$0.82$} & {$0.48\pm0.08$}\\
\vspace{0.13cm}
 {$B^+\rightarrow K^{*+}\omega$} &  {$0.64_{-0.06-0.02-0.03}^{+0.06+0.02+0.04}$} & {$0.67_{-0.04-0.39}^{+0.03+0.32}$} &  {$0.81$} &  {$0.41\pm0.19$}\\
\vspace{0.13cm}
 {$B^+\rightarrow K^{*+}\bar{K}^{*0}$} &  {$0.74_{-0.04-0.03-0.02}^{+0.03+0.02+0.01}$} & {$0.45_{-0.04-0.38}^{+0.02+0.55}$} &  {$0.75$} &  {$0.75\pm0.25$}\\
\vspace{0.3cm}
 {$B^+\rightarrow \rho^{+}\phi$} &  {$0.95_{-0.01-0.02-0.00}^{+0.01+0.01+0.00}$} & {$$} & {$$} & {$$}\\
 \hline\hline
\end{tabular}\label{tb:fl}
\footnotetext[1]{This is from BABAR data \cite{babar:rhorho}. The Belle's new measurement yields $0.21^{+0.18}_{-0.22} \pm0.13$ \cite{belle:rhorho}.}
\footnotetext[2]{This mode is employed as an input for extracting the parameters for $B \to K^*\rho$ decays in ref.\cite{qcdfbtovv}.}
\end{table}

\begin{table}[!t]
\centering
 \caption{Direct $CP$ asymmetries (\%) in the $B \to VV$ decays and comparison with the predictions from QCDF\cite{qcdfbtovv}. Experimental data are from the Particle Data Group\cite{pdg}. For $B^0\rightarrow K^{*0(+)}\rho^{0(-)}$, the data is from the ref.\cite{cpks+rho-}}
 \vspace{0.2cm}
\begin{tabular}[t]{lcccc}
\hline\hline
\vspace{0.25cm}
  {Decay Modes} &  { This work }&  {QCDF}& {Exp.}\\
 \hline
\vspace{0.13cm}
 {$B^0\rightarrow \rho^{0}\rho^{0}$}& {$70.7_{-5.2-5.4-6.0}^{+2.9+0.8+3.8}$} & {$30_{-16-26}^{+17+14}$}& {$$}\\
\vspace{0.13cm}
 {$B^0\rightarrow \rho^{+}\rho^{-}$}&  {$0.83_{-0.59-0.31-0.00}^{+0.50+0.66+0.00}$} & {$-4_{-0-3}^{+0+3}$} & {$$}\\
\vspace{0.13cm}
 {$B^0\rightarrow \rho^{0}\omega$}  &  {$59.4_{-8.3-5.5-6.3}^{+11.9+6.3+5.0}$} & {$3_{-6-76}^{+2+51}$} &  {$$}\\
\vspace{0.13cm}
 {$B^0\rightarrow \omega\omega$}  &  {$-73.7_{-6.2-6.0-0.9}^{+6.7+2.6+3.3}$} & {$-30_{-14-18}^{+15+16}$}&  {$$}\\
\vspace{0.13cm}
 {$B^0\rightarrow K^{*0}\rho^{0}$} &  {$-8.9_{-0.6-2.8-1.0}^{+0.6+2.8+1.1}$} & {$-15_{-8-14}^{+4+16}$} &  {$-6\pm9\pm2$}\\
\vspace{0.13cm}
 {$B^0\rightarrow K^{*+}\rho^{-}$} &  {$24.5_{-1.5-3.4-0.6}^{+1.2+2.9+0.0}$} & {$32_{-3-14}^{+1+2}$} &  {$21\pm15\pm2$}\\
\vspace{0.13cm}
 {$B^0\rightarrow K^{*0}\omega$} &  {$5.6_{-0.3-1.3-0.9}^{+0.3+1.2+0.8}$} & {$23_{-5-18}^{+9+5}$} & {$45\pm25$} \\
\vspace{0.13cm}
 {$B^0\rightarrow K^{*0}\bar{K}^{*0}$} &  {$0.0$} &  {$-14_{-1-2}^{+1+6}$} &  {$$} \\
\vspace{0.13cm}
 {$B^0\rightarrow K^{*+}K^{*-}$} &  {$29.8_{-5.7-9.5-4.7}^{+2.0+6.4+4.6}$} &  {$0$} & {$$} \\
\vspace{0.13cm}
 {$B^0\rightarrow \rho^{0}\phi$} &  {$0.0$} &  {$$} &  {$$} \\
\vspace{0.13cm}
 {$B^0\rightarrow \omega\phi$} &  {$0.0$} &  {$$} &  {$$} \\
\vspace{0.13cm}
 {$B^0\rightarrow \phi\phi$} &  {$0.0$} &  {$$} &  {$$} \\
\vspace{0.13cm}
 {$B^0\rightarrow K^{*0}\phi$} &  {$0.0$} &  {$0.8_{-0-0.5}^{+0+0.4}$} &  {$$} \\
\vspace{0.13cm}
 {$B^+\rightarrow \rho^{+}\rho^{0}$} &  {$0.05_{-0.01-0.03-0.00}^{-0.03+0.05+0.00}$} & {$0.06$} & {$-5\pm5$} \\
\vspace{0.13cm}
 {$B^+\rightarrow \rho^{+}\omega$} &  {$-11.2_{-2.0-2.5-0.6}^{+1.8+2.4+0.9}$} & {$-8_{-1-4}^{+1+3}$} &  {$-20\pm9$}\\
\vspace{0.13cm}
 {$B^+\rightarrow K^{*+}\rho^0$} &  {$22.7_{-1.5-2.5-1.2}^{+1.1+2.6+0.4}$} & {$43_{-2-28}^{+6+12}$} &  {$31\pm13$} \\
\vspace{0.13cm}
 {$B^+\rightarrow K^{*0}\rho^+$} &  {$-1.0_{-0.3-0.0-0.2}^{+0.2+0.2+0.1}$} & {$-0.3_{-0-0}^{+0+2}$} &  {$-1\pm16$} \\
\vspace{0.13cm}
 {$B^+\rightarrow K^{*+}\omega$} &  {$9.1_{-3.2-3.5-0.3}^{+3.3+1.3+0.0}$} & {$56_{-4-43}^{+3+4}$} & {$29\pm35$} \\
\vspace{0.13cm}
 {$B^+\rightarrow K^{*+}\bar{K}^{*0}$} &  {$23.0_{-4.2-2.2-1.4}^{+4.6+0.2+0.7}$} & {$16_{-3-34}^{+1+17}$} &  {$$} \\
\vspace{0.13cm}
 {$B^+\rightarrow K^{*+}\phi$} &  {$-1.0$} &  {$0.05$} &  {$-1\pm8$} \\
\vspace{0.3cm}
 {$B^+\rightarrow \rho^{+}\phi$} &  {$0.0$} &  {$$} &  {$$} \\
 \hline\hline
\end{tabular}\label{tb:cp}
\end{table}

\begin{table}[!t]
\centering
 \caption{Updated percentage of the transverse polarizations $f_{\perp}$(\%), relative phases $\phi_{\parallel}(rad)$,
$\phi_{\perp}(rad)$, $\Delta\phi_{\parallel}(10^{-2} rad)$, $\Delta\phi_{\perp}(10^{-2}rad)$ and the $CP$ asymmetry parameters $A^{0}_{CP}$(\%) and $A^{\perp}_{CP}$(\%) in $B\rightarrow VV$ decays calculated in the PQCD approach.}
 \vspace{0.2cm}
\begin{tabular}[t]{lccccccc}
\hline\hline
\vspace{0.3cm}
  {Decay Modes} &  {$f_{\perp}$ }&  {$\phi_{\parallel}$}& {$\phi_{\perp}$}& {$A^{0}_{CP}$}& {$A^{\perp}_{CP}$}& {$\Delta\phi_{\parallel}$}& {$\Delta\phi_{\perp}$}\\
 \hline
\vspace{0.13cm}
 {$B^0\rightarrow \rho^{0}\rho^{0}$}& {$45.9_{-8.2}^{+1.1}$} & {$2.68_{-1.09}^{+1.90}$} & {$2.81_{-1.95}^{+0.95}$}& {$88.9_{-120.7}^{+9.0}$}& {$-11.6_{-2.9}^{+16.2}$}& {$-98.9_{-69.6}^{+251.9}$}& {$-105_{-41}^{+266}$}\\
\vspace{0.13cm}
 {$B^0\rightarrow \rho^{+}\rho^{-}$}&  {$2.42_{-0.19}^{+0.21}$} & {$3.12_{-0.06}^{+0.06}$} & {$3.16_{-0.05}^{+0.06}$}& {$-2.05_{-0.55}^{+0.53}$}& {$39.0_{-8.4}^{+7.6}$}& {$10.2_{-3.1}^{+3.0}$}& {$9.58_{-3.19}^{+2.93}$}\\
\vspace{0.13cm}
 {$B^0\rightarrow \rho^{0}\omega$}  &  {$16.7_{-3.6}^{+5.0}$} & {$3.13_{-0.19}^{+0.17}$} & {$3.13_{-0.19}^{+0.17}$}& {$26.6_{-12.2}^{+19.8}$}& {$-60.0_{-12.1}^{+11.8}$}& {$-87.8_{-15.3}^{+13.7}$}& {$-98.4_{-15.1}^{+12.9}$}\\
\vspace{0.13cm}
 {$B^0\rightarrow \omega\omega$}  &  {$18.2_{-5.3}^{+6.1}$} & {$3.20_{-0.20}^{+0.25}$}& {$3.21_{-0.22}^{+0.24}$}& {$-5.70_{-16.2}^{+11.8}$}& {$17.0_{-22.1}^{+19.1}$}& {$105_{-10.4}^{+13.2}$}& {$108_{-11.1}^{+13.8}$}\\
\vspace{0.13cm}
 {$B^0\rightarrow K^{*0}\rho^{0}$} &  {$16.9_{-1.8}^{+2.7}$} & {$4.67_{-3.06}^{+0.02}$} & {$4.66_{-3.06}^{+0.01}$} & {$3.64_{-1.07}^{+1.20}$}& {$-7.71_{-1.86}^{+1.97}$}& {$-0.12_{-1.79}^{+1.72}$}& {$0.22_{-1.65}^{+1.85}$}\\
\vspace{0.13cm}
 {$B^0\rightarrow K^{*+}\rho^{-}$} &  {$15.6_{-2.5}^{+2.5}$} & {$3.31_{-0.21}^{+0.23}$} & {$3.30_{-0.21}^{+0.22}$} & {$23.8_{-5.1}^{+4.7}$}& {$-50.9_{-3.9}^{+4.9}$}& {$128_{-4.4}^{+4.1}$}& {$127_{-4.3}^{+43}$}\\
\vspace{0.13cm}
 {$B^0\rightarrow K^{*0}\omega$} &  {$18.3_{-2.3}^{+2.6}$} &  {$2.18_{-0.20}^{+0.21}$} & {$2.14_{-0.19}^{+0.21}$} & {$1.46_{-1.62}^{+1.44}$}& {$-8.92_{-4.01}^{+5.01}$}& {$-2.28_{-1.89}^{+1.79}$}& {$-12.0_{-4.9}^{+3.5}$}\\
\vspace{0.13cm}
 {$B^0\rightarrow K^{*0}\bar{K}^{*0}$} &  {$19.7_{-3.6}^{+4.0}$} & {$2.26_{-0.16}^{+0.20}$} & {$2.31_{-0.15}^{+0.19}$} & {$\sim0.0$}& {$\sim0.0$}& {$\sim0.0$}& {$\sim0.0$}\\
\vspace{0.13cm}
 {$B^0\rightarrow K^{*+}K^{*-}$} &  {$\sim0.0$} &  {$3.34_{-0.06}^{+0.08}$} & {$3.37_{-0.09}^{+2.60}$} & {$0.02_{-0.01}^{+0.02}$}& {$-75.3_{-10.5}^{+21.1}$}& {$56.4_{-9.7}^{+10.9}$}& {$-129_{-2.0}^{+258}$}\\
\vspace{0.13cm}
 {$B^0\rightarrow \rho^{0}\phi$} &  {$2.36_{-0.76}^{+1.08}$} & {$3.76_{-0.31}^{+0.22}$} & {$3.77_{-0.27}^{+0.24}$} & {$\sim0.0$}& {$\sim0.0$}& {$\sim0.0$}& {$\sim0.0$}\\
\vspace{0.13cm}
 {$B^0\rightarrow \omega\phi$} &  {$2.78_{-0.86}^{+1.08}$} &  {$3.77_{-0.28}^{+0.20}$} & {$3.78_{-0.25}^{+0.20}$} & {$\sim0.0$}& {$\sim0.0$}& {$\sim0.0$}& {$\sim0.0$}\\
\vspace{0.13cm}
 {$B^0\rightarrow \phi\phi$} &  {$0.05_{-0.02}^{+0.02}$} &  {$3.26_{-0.14}^{+0.20}$} & {$3.50_{-0.17}^{+0.17}$} & {$\sim0.0$}& {$\sim0.0$}& {$\sim0.0$}& {$\sim0.0$}\\
\vspace{0.13cm}
 {$B^+\rightarrow \rho^{+}\rho^{0}$} &  {$0.46_{-0.06}^{+0.08}$} & {$3.20_{-0.09}^{+0.07}$} & {$3.18_{-0.10}^{+0.07}$} & {$0.002_{-0.003}^{+0.003}$}& {$-0.32_{-0.64}^{+0.25}$}& {$-0.11_{-0.32}^{+0.10}$}& {$-0.79_{-0.45}^{+0.11}$}\\
\vspace{0.13cm}
 {$B^+\rightarrow \rho^{+}\omega$} &  {$1.18_{-0.29}^{+0.38}$} & {$2.64_{-0.15}^{+0.14}$} & {$2.57_{-0.15}^{+0.16}$} & {$-2.02_{-0.74}^{+0.69}$}& {$76.2_{-14.7}^{+11.0}$}& {$70.7_{-16.1}^{+16.8}$}& {$83.9_{-19.7}^{+17.3}$}\\
\vspace{0.13cm}
 {$B^+\rightarrow K^{*+}\rho^0$} &  {$11.9_{-2.0}^{+2.3}$} &  {$1.94_{-0.14}^{+1.44}$} & {$1.94_{-0.15}^{+1.43}$} & {$11.3_{-2.4}^{+2.3}$}& {$-34.0_{-2.8}^{+3.7}$}& {$-26.4_{-4.0}^{+157}$}& {$-27.3_{-4.0}^{+158}$}\\
\vspace{0.13cm}
 {$B^+\rightarrow K^{*0}\rho^+$} &  {$13.7_{-1.9}^{+2.1}$} &  {$1.81_{-0.18}^{+0.20}$} & {$1.81_{-0.18}^{+0.19}$} & {$-0.36_{-0.11}^{+0.12}$}& {$0.98_{-0.25}^{+0.20}$}& {$-1.19_{-0.36}^{+0.38}$}& {$-1.54_{-0.49}^{+0.41}$}\\
\vspace{0.13cm}
 {$B^+\rightarrow K^{*+}\omega$} & {$17.2_{-3.5}^{+3.4}$}& {$2.18_{-0.20}^{+0.23}$} & {$2.18_{-0.20}^{+0.22}$}&  {$11.2_{-4.3}^{+3.9}$} &  {$-19.9_{-3.6}^{+5.5}$} & {$-37.9_{-6.1}^{+7.0}$}& {$-38.7_{-6.1}^{+7.2}$}\\
\vspace{0.13cm}
 {$B^+\rightarrow K^{*+}\bar{K}^{*0}$} &  {$12.9_{-2.4}^{+1.7}$} & {$1.98_{-0.17}^{+0.20}$} & {$1.99_{-0.19}^{+0.18}$} & {$7.21_{-2.50}^{+2.54}$}& {$-19.1_{-2.6}^{+4.2}$}& {$20.2_{-5.8}^{+4.8}$}& {$28.4_{-6.2}^{-7.7}$}\\
\vspace{0.3cm}
 {$B^+\rightarrow \rho^{+}\phi$} &  {$2.36_{-0.76}^{+1.08}$} & {$3.76_{-0.31}^{+0.22}$} & {$3.77_{-0.27}^{+0.23}$} & {$\sim0.0$}& {$\sim0.0$}& {$\sim0.0$}& {$\sim0.0$}\\
 \hline\hline
\end{tabular}\label{tb:ft}
\end{table}

\begin{table}[!t]
\centering
 \caption{Updated branching ratios (in units of $10^{-6}$) of $B_s\rightarrow VV$ decays calculated in the PQCD approach. For comparison, we also cite the updated theoretical predictions in the QCDF approach~\cite{qcdfbsvv} and the previous predictions in the PQCD approach~\cite{bsvv}.
 Experimental data are from the Particle Data Group~\cite{pdg}}
 \vspace{0.1cm}
\begin{tabular}[t]{lccccc}
\hline\hline
\vspace{0.3cm}
 {Decay Modes}&  {Class} &  {This work}& { QCDF}& { PQCD(former)}&  {Exp}\\
 \hline
\vspace{0.2cm}
 {$B_s\rightarrow K^{*-}\rho^{+}$}& {T} & {$24.0_{-8.7-1.4-2.4}^{+10.9+1.2+0.0}$}& {$21.6_{-2.8-1.5}^{+1.3+0.9}$}& {$20.9_{-6.2-1.4-1.1}^{+8.2+1.4+1.2}$}& {$$} \\
\vspace{0.2cm}
 {$B_s\rightarrow \bar{K}^{*0}\rho^{0}$}&  {C}& {$0.40_{-0.15-0.07-0.03}^{+0.19+0.11+0.00}$}& {$1.3_{-0.6-0.3}^{+2.0+1.7}$}& {$0.33_{-0.07-0.09-0.01}^{+0.09+0.14+0.00}$}& {$<767$} \\
\vspace{0.2cm}
 {$B_s\rightarrow \bar{K}^{*0}\omega$} &  {C} & {$0.35_{-0.14-0.08-0.08}^{+0.16+0.09+0.04}$} & {$1.1_{-0.5-0.3}^{+1.5+1.3}$}& {$0.31_{-0.07-0.06-0.02}^{+0.10+0.12+0.04}$}& {$$} \\
\vspace{0.2cm}
 {$B_s\rightarrow K^{*+}K^{*-}$} &  {P} &  {$5.4_{-1.7-1.4-0.5}^{+2.7+1.8+0.3}$} & {$7.6_{-1.0-1.8}^{+1.0+2.3}$}& {$6.7_{-1.2-1.4-0.2}^{+1.5+3.4+0.5}$}& {$$} \\
\vspace{0.2cm}
 {$B_s\rightarrow \rho^{0}\phi$} &  {P} & {$0.23_{-0.05-0.01-0.02}^{+0.15+0.03+0.01}$} & {$0.18_{-0.01-0.04}^{+0.01+0.09}$}& {$0.23_{-0.07-0.01-0.01}^{+0.09+0.03+0.00}$}& {$<617$} \\
\vspace{0.2cm}
 {$B_s\rightarrow \omega\phi$} &  {P} &  {$0.17_{-0.07-0.04-0.01}^{+0.10+0.05+0.00}$} & {$0.18_{-0.12-0.04}^{+0.44+0.47}$}& {$0.16_{-0.05-0.04-0.00}^{+0.09+0.10+0.01}$}& {$$} \\
\vspace{0.2cm}
 {$B_s\rightarrow \rho^+\rho^{-}$} &  {P} &  {$1.5_{-0.6-0.2-0.1}^{+0.7+0.2+0.0}$} & {$0.68_{-0.04-0.53}^{+0.04+0.73}$}& {$1.0_{-0.2-0.2-0.0}^{+0.2+0.3+0.0}$}& {$$} \\
\vspace{0.2cm}
 {$B_s\rightarrow \rho^0\rho^0$} &  {P} & {$0.74_{-0.24-0.14-0.00}^{+0.39+0.22+0.00}$} & {$0.34_{-0.02-0.26}^{+0.02+0.36}$}& {$0.51_{-0.11-0.10-0.01}^{+0.12+0.17+0.01}$}& {$<320$} \\
\vspace{0.2cm}
 {$B_s\rightarrow \rho^0\omega$} &  {E} & {$0.009_{-0.003-0.002-0.001}^{+0.003+0.001+0.000}$} & {$0.004_{-0.0-0.003}^{+0.0+0.005}$}& {$0.007_{-0.001-0.001-0.000}^{+0.002+0.001+0.000}$}& {$$} \\
\vspace{0.4cm}
 {$B_s\rightarrow \omega\omega$} &  {P} & {$0.40_{-0.18-0.10-0.01}^{+0.16+0.10+0.00}$} & {$0.19_{-0.02-0.15}^{+0.02+0.21}$} & {$0.39_{-0.08-0.07-0.00}^{+0.09+0.13+0.01}$}& {$$}\\
 \hline\hline
\end{tabular}\label{tb:brbs}
\end{table}

\begin{table}[!t]
\centering
 \caption{Percentage of the longitudinal polarizations $f_{L}$ in $B_s \to VV $ decays
and comparison with the QCDF approach~\cite{qcdfbsvv} and the previous predictions in the PQCD approach~\cite{bsvv}.}
 \vspace{0.1cm}
\begin{tabular}[t]{lccc}
\hline\hline
\vspace{0.3cm}
 {Decay Modes} &  {This work}& { QCDF}& { PQCD(former)}\\
 \hline
\vspace{0.2cm}
 {$B_s\rightarrow K^{*-}\rho^{+}$}& {$0.95_{-0.01-0.01-0.00}^{+0.01+0.01+0.00}$}& {$0.92_{-0.02-0.03}^{+0.01+0.01}$}& {$0.937_{-0.002-0.003-0.002}^{+0.001+0.002+0.000}$} \\
\vspace{0.2cm}
 {$B_s\rightarrow \bar{K}^{*0}\rho^{0}$}& {$0.57_{-0.10-0.08-0.00}^{+0.06+0.06+0.01}$}& {$0.90_{-0.05-0.23}^{+0.04+0.03}$}& {$0.455_{-0.003-0.043-0.009}^{+0.004+0.069+0.006}$} \\
\vspace{0.2cm}
 {$B_s\rightarrow \bar{K}^{*0}\omega$} &  {$0.50_{-0.08-0.15-0.01}^{+0.07+0.11+0.01}$} & {$0.90_{-0.04-0.23}^{+0.03+0.03}$}& {$0.532_{-0.002-0.029-0.013}^{+0.003+0.035+0.023}$} \\
\vspace{0.2cm}
 {$B_s\rightarrow K^{*+}K^{*-}$}&  {$0.42_{-0.09-0.03-0.06}^{+0.13+0.03+0.05}$} & {$0.52_{-0.05-0.21}^{+0.03+0.20}$}& {$0.438_{-0.040-0.023-0.015}^{+0.051+0.021+0.037}$} \\
\vspace{0.2cm}
 {$B_s\rightarrow \rho^{0}\phi$} &  {$0.86_{-0.01-0.01-0.00}^{+0.01+0.01+0.00}$} & {$0.88_{-0.00-0.18}^{+0.01+0.02}$}& {$0.870_{-0.002-0.003-0.004}^{+0.002+0.009+0.009}$} \\
\vspace{0.2cm}
 {$B_s\rightarrow \omega\phi$}  &  {$0.69_{-0.09-0.09-0.02}^{+0.08+0.08+0.02}$} & {$0.95_{-0.02-0.42}^{+0.01+0.00}$}& {$0.443_{-0.075-0.061-0.004}^{+0.000+0.054+0.009}$}\\
\vspace{0.2cm}
 {$B_s\rightarrow \rho^+\rho^{-}$} &  {$\sim1.0$} & {$\sim1.0$}& {$\sim1.0$} \\
\vspace{0.2cm}
 {$B_s\rightarrow \rho^0\rho^0$} &  {$\sim1.0$} & {$\sim1.0$}& {$\sim1.0$} \\
\vspace{0.2cm}
 {$B_s\rightarrow \rho^0\omega$}  &  {$\sim1.0$} & {$\sim1.0$}& {$\sim1.0$} \\
\vspace{0.4cm}
 {$B_s\rightarrow \omega\omega$} &  {$\sim1.0$} & {$\sim1.0$} & {$\sim1.0$}\\
 \hline\hline
\end{tabular}\label{tb:flbs}
\end{table}

\begin{table}[!t]
\centering
 \caption{Direct $CP$ asymmetries (\%) in the $B_s\to VV$ decays and comparison with the QCDF approach~\cite{qcdfbsvv}
 and the previous predictions in the PQCD approach~\cite{bsvv}.}
 \vspace{0.1cm}
\begin{tabular}[t]{lccc}
\hline\hline
\vspace{0.3cm}
 {Decay Modes}&  {This work}& { QCDF}& { PQCD(former)}\\
 \hline
\vspace{0.2cm}
 {$B_s\rightarrow K^{*-}\rho^{+}$} & {$-9.1_{-1.5-1.2-0.3}^{+1.4+1.0+0.2}$}& {$-11_{-1-1}^{+1+4}$}& {$-8.2_{-1.2-1.7-1.1}^{+1.0+1.2+0.4}$} \\
\vspace{0.2cm}
 {$B_s\rightarrow \bar{K}^{*0}\rho^{0}$}& {$62.7_{-5.9-16.0-7.9}^{+6.4+10.5+7.5}$}& {$46_{-17-25}^{+15+10}$}& {$61.8_{-4.7-22.8-2.3}^{+3.2+17.1+4.4}$} \\
\vspace{0.2cm}
 {$B_s\rightarrow \bar{K}^{*0}\omega$}&  {$-78.1_{-2.2-7.4-8.3}^{+2.9+13.1+8.1}$} & {$-50_{-15-6}^{+20+21}$}& {$-62.1_{-3.9-12.6-1.9}^{+4.8+19.7+5.5}$} \\
\vspace{0.2cm}
 {$B_s\rightarrow K^{*+}K^{*-}$} &  {$8.8_{-8.9-2.9-0.2}^{+2.5+0.5+0.0}$} & {$21_{-2-4}^{+1+2}$}& {$9.3_{-0.7-3.6-0.2}^{+0.4+3.3+0.3}$} \\
\vspace{0.2cm}
 {$B_s\rightarrow \rho^{0}\phi$} &  {$-4.3_{-0.5-0.5-1.0}^{+0.6+0.6+1.2}$} & {$83_{-0.0-36}^{+1.0+10}$}& {$10.1_{-0.9-1.8-0.5}^{+0.9+1.6+1.3}$} \\
\vspace{0.2cm}
 {$B_s\rightarrow \omega\phi$} &  {$28.0_{-3.2-2.3-5.1}^{+1.3+0.5+3.4}$} & {$-8_{-1-15}^{+3+20}$}& {$3.6_{-0.6-2.4-0.2}^{+0.6+2.4+0.6}$} \\
\vspace{0.2cm}
 {$B_s\rightarrow \phi\phi$}  &  {$0.0$} & {$0.2_{-0.3-0.2}^{+0.4+0.5}$}& {$0.0$}\\
\vspace{0.2cm}
 {$B_s\rightarrow \rho^+\rho^{-}$} &  {$-2.9_{-1.1-1.3-0.2}^{+0.7+1.5+0.2}$} & {$0$}& {$-2.1_{-0.1-1.3-0.1}^{+0.2+1.7+0.1}$} \\
\vspace{0.2cm}
 {$B_s\rightarrow \rho^0\rho^0$} &  {$-2.9_{-1.1-1.3-0.2}^{+0.7+1.5+0.2}$} & {$0$}& {$-2.1_{-0.1-1.3-0.1}^{+0.2+1.7+0.1}$}\\
\vspace{0.2cm}
 {$B_s\rightarrow \rho^0\omega$}  &  {$11.1_{-1.5-4.4-1.4}^{+1.0+1.9+1.2}$} & {$0$}& {$6.0_{-0.5-3.9-0.4}^{+0.7+2.7+1.0}$}\\
\vspace{0.4cm}
 {$B_s\rightarrow \omega\omega$} &  {$-3.3_{-1.0-1.4-0.2}^{+0.8+1.5+0.5}$} & {$0$} & {$-2.0_{-0.1-1.3-0.1}^{+0.1+1.7+0.1}$}\\
 \hline\hline
\end{tabular}\label{tb:cpbs}
\end{table}

\begin{table}[!t]
\centering
 \caption{Updated percentage of the transverse polarizations $f_{\perp}$(\%), relative phases
 $\phi_{\parallel}({\rm rad})$, $\phi_{\perp}({\rm rad})$, $\Delta\phi_{\parallel}(10^{-2} {\rm rad})$,
$\Delta\phi_{\perp}(10^{-2} {\rm rad})$ and the $CP$ asymmetry parameters $A^{0}_{CP}$ and $A^{\perp}_{CP}$ in $B_s\rightarrow VV$ decays calculated in the PQCD approach.}
 \vspace{0.1cm}
\begin{tabular}[t]{lccccccc}
\hline\hline
\vspace{0.3cm}
 {Decay Modes} & {$f_{\perp}$}& {$\phi_{\parallel}$}& {$\phi_{\perp}$}& {$A^{0}_{CP}$}&
 {$A^{\perp}_{CP}$}& {$\Delta\phi_{\parallel}$}& {$\Delta\phi_{\perp}$}\\
 \hline
\vspace{0.2cm}
 {$B_s\rightarrow K^{*-}\rho^{+}$}& {$2.31_{-0.21}^{+0.22}$}& {$3.07_{-0.09}^{+0.07}$}& {$3.07_{-0.08}^{+0.08}$}& {$-2.71_{-0.72}^{+0.68}$}& {$55.0_{-10.5}^{+10.3}$}
& {$12.4_{-4.7}^{+4.8}$}& {$12.5_{-4.8}^{+4.5}$} \\
\vspace{0.2cm}
 {$B_s\rightarrow \bar{K}^{*0}\rho^{0}$}& {$22.5_{-4.7}^{+7.3}$}& {$1.94_{-0.10}^{+2.52}$}& {$1.99_{-0.10}^{+2.53}$}& {$-17.5_{-13.0}^{+21.2}$}& {$22.0_{-31.4}^{+29.9}$}&
{$-31.5_{-16.2}^{+274}$}& {$-36.5_{-15.8}^{+222}$}\\
\vspace{0.2cm}
 {$B_s\rightarrow \bar{K}^{*0}\omega$} &  {$26.1_{-7.0}^{+9.8}$} & {$2.18_{-0.28}^{+0.33}$}& {$2.23_{-0.27}^{+0.32}$}& {$-5.99_{-50.21}^{+23.52}$}& {$6.95_{-32.14}^{+27.91}$}& {$30.7_{-24.3}^{+30.9}$}
& {$36.5_{-24.2}^{+31.3}$}\\
\vspace{0.2cm}
 {$B_s\rightarrow K^{*+}K^{*-}$}&  {$27.7_{-7.0}^{+5.2}$} & {$3.53_{-0.25}^{+0.33}$}& {$3.54_{-0.24}^{+0.36}$}& {$45.4_{-23.4}^{+19.0}$}& {$-32.9_{-4.0}^{+5.6}$}& {$93.7_{-14.1}^{+11.1}$}
& {$93.4_{-13.8}^{+11.1}$} \\
\vspace{0.2cm}
 {$B_s\rightarrow \rho^{0}\phi$} &  {$8.89_{-1.06}^{+0.80}$} & {$3.11_{-0.09}^{+0.10}$}& {$3.29_{-0.09}^{+0.09}$}& {$3.27_{-1.19}^{+1.07}$}& {$-32.8_{-5.8}^{+7.4}$}&
 {$-43.7_{-9.5}^{+9.9}$}& {$-63.9_{-9.6}^{+10.8}$} \\
\vspace{0.2cm}
 {$B_s\rightarrow \omega\phi$}  &  {$16.1_{-5.8}^{+7.3}$} & {$3.38_{-0.17}^{+0.20}$}& {$3.35_{-0.23}^{+0.30}$}& {$-2.24_{-5.45}^{+6.67}$}& {$4.38_{-15.93}^{+17.52}$}& {$-36.7_{-11.9}^{+12.5}$}
& {$-32.7_{-18.6}^{+16.5}$}\\
\vspace{0.2cm}
 {$B_s\rightarrow \rho^+\rho^{-}$} &  {$\sim0.0$} & {$3.40_{-0.04}^{+0.04}$}& {$3.27_{-0.15}^{+0.16}$}& {$0.0$}& {$30.5_{-16.3}^{+15.0}$}& {$2.87_{-0.59}^{+0.44}$}& {$-27.4_{-6.0}^{+6.9}$} \\
\vspace{0.2cm}
 {$B_s\rightarrow \rho^0\rho^0$} &  {$\sim0.0$} & {$3.40_{-0.04}^{+0.04}$}& {$3.27_{-0.15}^{+0.16}$}& {$0.0$}& {$30.5_{-16.3}^{+15.0}$}& {$2.87_{-0.59}^{+0.44}$}& {$-27.4_{-6.0}^{+6.9}$} \\
\vspace{0.2cm}
 {$B_s\rightarrow \rho^0\omega$}  &  {$\sim0.0$} & {$3.48_{-0.05}^{+0.04}$}& {$2.63_{-0.22}^{+0.18}$}& {$0.0$}& {$27.9_{-9.9}^{+9.3}$}& {$-9.30_{-5.23}^{+1.50}$}& {$-30.4_{-23.4}^{+19.1}$}     \\
\vspace{0.4cm}
 {$B_s\rightarrow \omega\omega$} &  {$\sim0.0$} & {$3.40_{-0.04}^{+0.04}$} & {$3.27_{-0.11}^{+0.16}$}& {$0.0$}& {$30.8_{-15.3}^{+14.0}$}& {$2.71_{-0.52}^{+0.42}$}& {$-26.7_{-5.7}^{+6.3}$}\\
 \hline\hline
\end{tabular}\label{tb:ftbs}
\end{table}

For the charmless $B_{q} \to VV$ decays, it is naively expected that the helicity amplitudes $H_{i}$ (with helicity $i=0,-,+$) satisfy the hierarchy pattern
\begin{eqnarray}
H_0:H_-:H_+=1: \frac{\Lambda_{QCD}}{m_b}:(\frac{\Lambda_{QCD}}{m_b})^2.
\label{eq:hierarchy}
\end{eqnarray}
In the naive factorization approach, longitudinal polarizations dominate the branching ratios of $B$ decays. In sharp contrast to these expectations, large transverse polarization of order 50$\%$ is observed in $B\to K^*\phi$, $B\to K^*\rho$ and  $B_s\to \phi\phi$ decays, which poses an interesting challenge for the theory. This shows that the scaling behavior shown in Eq.~(\ref{eq:hierarchy}) is violated. In order to interpret this large transverse polarization many mechanisms have been proposed, such as the penguin-induced annihilation contributions~\cite{flpenguin}, final-state interactions~\cite{flfsi}, form-factor tuning~\cite{flfft}, and even onset of new physics~\cite{flnp}.

As pointed out in the context of QCDF~\cite{qcdf3}, after taking into account the NLO effects, e.g., vertex-, penguin- and hard spectator-scattering contributions, the effective Wilson coefficients $a_i^h$ become helicity dependent. Including these effects, for some penguin-dominant modes, the constructive (destructive) interference in the transverse (longitudinal) amplitudes of $B \to VV$ decays makes the total transverse contribution comparable to the longitudinal one, and the transverse polarization fraction may reach as high as $50\%$.

In order to interpret the observed large transverse polarization fraction in the penguin-dominated $B\to VV$ decays, e.g., $B\to K^*\phi$, $B\to K^*\rho$, both the PQCD and the QCDF frameworks rely on penguin annihilation. However, in QCDF, the penguin-annihilation amplitude involves a troublesome endpoint divergence, which is fudged by introducing non-perturbative parameters. Hence, in QCDF, one can fit the existing experimental data on the branching ratios, $f_L$ and the $CP$ asymmetries by adjusting the annihilation parameters $\rho_A$ and $\phi_A$, which reduces the predictive power of the theory. In contrast, in the PQCD approach, the annihilation type diagrams can be perturbatively calculated without introducing any fudge factor (or parameter), which allows us to predict the direct $CP$ asymmetry and transverse polarization. The large transverse polarization fraction can be interpreted on the basis of the chirally enhanced annihilation diagrams, especially the $(S-P)(S+P)$ penguin annihilation, introduced by the QCD penguin operator  $O_6$ \cite{flpqcd}. A nice feature of the $(S-P)(S+P)$ penguin annihilation operator is that the light quarks in the final states are not produced through chiral currents. So, there is no suppression caused by the helicity flip. As a result,  the polarization fractions satisfy
\begin{eqnarray}
f_L\approx f_{\parallel} \approx f_{\perp}.
\end{eqnarray}
Thus, in the PQCD approach, the penguin-annihilation together with the hard-scattering emission diagrams can explain the large transverse polarization fraction measured in experiments.

We present our numerical results for the branching ratios, direct $CP$ asymmetries, and some other observables introduced earlier in the text, in Tables \ref{tb:zui}-\ref{tb:ftbs}. The dominant topologies contributing to these decays are also indicated in the tables through the symbols $T$ (the color-allowed tree contributions), $C$ (the color-suppressed tree contributions), $P$ (penguin contributions), and $E$ ($W$-exchange annihilation contributions). Theoretical uncertainties quoted in the tables are estimated from three sources: the first error quoted is from the input hadronic parameters, such as the decay constants of the initial $B_q$ and the final vector-mesons and the parameters in the distribution amplitudes of the initial and final states, which can be found in sec.~\ref{sec:function} and Eq.~(\ref{para}). The second error arises from the scale uncertainties, characterized by $\Lambda_{QCD}=(0.25\pm0.05)$ GeV and the variations of the factorization scales $t$ (from $0.8t$ to $1.2t$) detailed in Appendix \ref{app:a}. The scale-dependent uncertainty can be reduced only if the next-to-leading order contributions are known. The last error is the combined uncertainty in the CKM matrix elements and the angles of the unitary triangle. In Tables~\ref{tb:zui},\ref{tb:ft} and \ref{tb:ftbs}, we have combined these uncertainties by adding them in quadrature and show the resulting uncertainty, due to to the space limitations in the Tables.

We now discuss these results. For the branching ratios, the most important theoretical uncertainty is the first error caused by the nonperturbative input parameters. In the PQCD approach, the wave functions are the primary important input parameters and they heavily influence the predictions of the branching ratios, as also discussed in~\cite{xiaowang}. We have adopted the new updated wave functions.  While, for the direct $CP$ asymmetry parameters, the dominant uncertainty arises from the second error, which is caused by the unknown higher order QCD corrections. From the definition:
\begin{eqnarray}
A_{CP}^{dir}&\equiv&\frac{BR(\bar{B}\to f)-BR(B\to \bar{f})}{BR(\bar{B}\to f)+BR(B\to \bar{f})}\nonumber\\
&=&\frac{\mid A(\bar{B}\to f)\mid^2-\mid A(B\to \bar{f})\mid^2}{\mid A(\bar{B}\to f)\mid^2+\mid A(B\to \bar{f})\mid^2},
\end{eqnarray}
it is apparent that the wave functions of the initial $B_{q}$ meson and the final vector mesons are overall factors, hence they drop out in the ratio and do not provide significant contributions in the estimates of the direct $CP$ asymmetries.  Direct $CP$ asymmetries are proportional to the strong phases originated from the hard part, and the NLO QCD corrections will influence the strong phases significantly. Not having these corrections at our disposal, we can only estimate them by varying the scales. The resulting theoretical uncertainty is larger than the one from the wave functions, and we assume that the variation of the scales is an adequate account of the NLO corrections at this stage.

For comparison, the updated results of the QCDF approach\cite{qcdfbtovv,qcdfbsvv} and the earlier PQCD predictions\cite{bsvv,btovv,jpg32} are also presented. We have updated the PQCD computations in this work and the main improvements are: (i) Use of the updated vector mesons distribution amplitudes with new estimates of the Gegenbauer moments and decay constants, and (ii) the treatment of the terms in the decay amplitude proportional to the ratio $r_{i}^2=m_{V_i}^2/m_B^2(i=2,3)$. Since wave functions are the most important inputs in PQCD, their improved knowledge is expected to yield improved estimates of the branching ratios, polarization fractions, and other observables. We recall that in the earlier PQCD computations,  $r_{i}^2$-dependent terms in the denominator of the propagators of the virtual quarks and gluons were omitted. From Appendix A, we find that, although their contribution is formally power suppressed, it can numerically change the real and imaginary parts of amplitudes and enhance the transverse polarization component, especially for the penguin-dominant decays. To quantify this, we have listed the amplitudes, branching ratios and transverse polarization fractions  of the penguin-dominant  decays $B^0\to K^{*0}\phi$,  $B_s \to \phi\phi$ and the tree-dominant decay $B^+\to \rho^+\rho^0$ with and without the  $r_i^2$-terms in Table\ref{tb:amp}. We note that for the two penguin-dominant decays, the impact of the $r_i^2$-dependent terms in the amplitudes of the annihilation part, as well as in the imaginary part of the emission diagrams, is numerically significant. Taking the factorizable annihilation diagrams as an example, in the range near $x_3 \to 1$ or $x_2 \to 0$, the nonzero $r_i^2$ contributes a non-negligible imaginary part. So by keeping the $r_i^2$-terms, the branching ratios are reduced, while the transverse polarization fractions rise.  The two main improvements go in the right direction in explaining the observed branching ratios and the large transverse polarization fractions in $B \to K^*\phi$ and $B_s\to \phi\phi$ decays in the PQCD approach. For the tree-dominant decay $B^+\to \rho^+\rho^0$, however, the effect on the traditional emission diagrams produced by the $r_i^2$-terms is tiny, as expected. This is further discussed in Appendix A. Thus, the improved PQCD treatment presented here yields better consistency with the data.

\begin{table}[!t]
\centering
 \caption{Amplitudes ($10^{-3}$), branching ratios ($10^{-6}$) and the polarization fractions ($\%$) with (and  without)
the $r_i^2$-dependent terms in the $B^0\to K^{*0}\phi$, $B_s \to \phi\phi$ and $B^0\to \rho^+\rho^0$ decays. }
\begin{tabular}[t]{lcccccc}
\hline\hline
 Modes& & $A^{L}$ &$A^{N}$ &$A^T$& Br& $f_{L}$\\
 \hline
{$B^0\rightarrow K^{*0}\phi(r_i^2)$}&emission &-3.3+0.67$i$ &-0.66+0.06$i$&0.64-0.05$i$&{9.8}&{56}\\
&annihilation &0.32-1.6$i$ &-0.43+0.84$i$&0.42-0.83$i$&&\\
{$B^0\rightarrow K^{*0}\phi$}&emission& -3.0-0.09$i$ &-0.71-0.012$i$&0.69+0.03$i$&{15}&{70}\\
& annihilation& -0.42-1.95i &0.05+1.28$i$&-0.11-1.38$i$& &\\
\hline
{$B_s\rightarrow \phi\phi(r_i^2)$}&emission &-2.8+0.37$i$ &-0.60+0.10$i$&0.60-0.08$i$&{16.7}&{34.7}\\
&annihilation &0.68-1.2$i$ &-0.53+1.0$i$&0.53-1.0$i$&&\\
{$B_s\rightarrow \phi\phi$}&emission& -2.6-0.02$i$ &-0.64+0.03$i$&0.63-0.005$i$&{26.6}&{45}\\
& annihilation& -0.04-1.8$i$ &0.18+1.8$i$&-0.15-1.7$i$& &\\
 \hline
{$B^+\rightarrow \rho^+\rho^0(r_i^2)$}&emission &3.0+5.9$i$&0.28+0.33$i$&0.27-0.29$i$&{13.5}&{98}\\
&annihilation &$\sim0$ &$\sim0$&$\sim0$&&\\
{$B^+\rightarrow \rho^+\rho^0$}&emission& 2.8+5.8$i$ &0.12+0.33$i$&-0.11-0.29$i$&{13.3}&{99}\\
& annihilation& $\sim0$ &$\sim0$&$\sim0$& &\\
 \hline\hline
\end{tabular}\label{tb:amp}
\end{table}

In Table~\ref{tb:zui}, we list the current experimental measurements in $B^0(B^+) \to K^* (K^{*+}) \phi$ and $B_s \to \overline K^{*0}\phi$, $B_s \to \overline K^{*0}K^{*0}$ and $B_s \to \phi \phi$ decays and compare them with our theoretical results worked out in this paper. These decays are all penguin-dominated and are measured with large fraction of transverse polarization. For these decays, the naive factorization approach  predicts too small branching ratios by a factor of $2\sim 3$ \cite{qcdfbtovv}, due to the small contribution from the penguin operators. In~\cite{btovv,bsvv,jpg32,flfft}, the authors have studied these $B_q\to VV$ decays, but those predictions are not in good agreement with the currently available experimental data. The primary task is to bring up the branching ratios and explain the polarization anomaly in these decays. In our update, we explain the bulk of the data. However, we note that for $B_s \to K^{*0}\overline{K}^{*0}$ and $B_s \to \overline{K}^{*0}\phi$ modes, our calculated branching ratios are $(5.4_{-2.4}^{+3.0})\times 10^{-6}$ and $(0.39_{-0.19}^{+0.20})\times 10^{-6}$ respectively, which are much smaller than the data, though they are compatible with the QCDF predictions $(6.6_{-1.4-1.7}^{+1.1+1.9})\times 10^{-6}$ and $(0.37_{-0.05-0.20}^{+0.06+0.24})\times10^{-6}$ respectively.

In Table~\ref{tb:br}, we have given our estimates of the $B \to VV$ branching ratios for different topologies. For the penguin dominant decay modes (indicated by $P$ in the tables), our updated predictions basically  agree with the QCDF predictions, except for $B^0\to K^{*0}\omega$. Due to the constructive interference between the penguin emission contributions and the penguin annihilation contributions, our prediction for this decays is almost twice as large as that of QCDF, and it also comes out larger than the current experimental data. As the experimental error is still large, we wait for consolidated date from Belle-II experiment. For the color-suppressed decay $B^0 \to\rho^0\rho^0$, the calculated branching fraction in this work is $(0.27_{-0.09-0.04-0.01}^{+0.10+0.06+0.00})\times 10^{-6}$, while BaBar and Belle obtained $(0.9\pm 0.32\pm0.14)\times10^{-6}$ \cite{babar} and $(0.4\pm0.4_{-0.3}^{+0.2})\times10^{-6}$ \cite{belle}, respectively, with the current world average being $(0.73\pm0.28)\times 10^{-6}$. Our result, within errors, agrees with the Belle data. Judged from the isospin triangle, since the decay rate of $B^0\to \rho^0\rho^0$ is so small, the rate for the decay $B^0 \to \rho^+\rho^-$ ought to be double that of $B^+\to \rho^+\rho^0$. In experiment, however, within errors, these two rates are equal to each other, which is puzzling. Thus, the experimental situation is still in a state of flux. In Table \ref{tb:fl}, discussed in more detail later, we show that for the $B^0\to \rho^0\rho^0$ decay the longitudinal polarization fraction is as small as $12\%$. As is well known, for the  color-suppressed decays, the longitudinal polarization contributions from two hard-scattering emission diagrams largely cancel against each other. What's worse, the remaining longitudinal polarization contributions are nearly canceled by those from the annihilation diagrams.  On the other hand, the chiral enhanced annihilation diagrams and hard-scattering emission diagrams provide a large transverse polarizable contribution. In the end, the $B^0\to \rho^0\rho^0$ is almost totally dominated by the transverse polarization component. In Table~\ref{tb:fl}, we adopt the BaBar data \cite{babar:rhorho}, but note that Belle has provided a new measurement $0.21^{+0.18}_{-0.22}\pm{0.13}$~\cite{belle:rhorho}, which supports our theoretical calculations. Thus, it is important to have a refined measurement of the branching fractions and the longitudinal polarization fractions for  $B\to \rho\rho$ to draw definitive conclusion. It should be noted that if the next leading order corrections are included, the branching fraction of $B\to \rho^0\rho^0$ might be enlarged while its transverse polarization fraction $f_\perp$ will become smaller \cite{Li:2006cva}. The previous PQCD estimates for the $B^0 \to \rho^0 \omega$ decay rate exceeded the current experimental upper bound. In this work, this branching ratio is now lower than the upper experimental bound but is about a factor five larger than the QCDF prediction due to a near cancelation of the color-suppressed tree amplitudes. In the framework of PQCD, although the color-suppressed tree amplitudes also almost mutually cancel, the decay can get significant contributions from the annihilation type diagrams so that the decay rate comes back up and is not as small as in the QCDF prediction. This, together with some other predictions, provides an experimental check on these two competing frameworks.

In Table.~\ref{tb:fl}, we have given the fraction of the longitudinal polarization component, $f_L$ for $B \to VV$ decays, where we have compared them with the available data, and also with the previous PQCD~\cite{btovv} and QCDF~\cite{qcdfbtovv} approaches. Of these, the predictions for the decays $B\to \phi\rho(\omega)$ are worked out for the first time. For the $B^0\to \rho^0\omega$ decay, we predict the longitudinal polarization fraction as small as $67\%$, which is due to a significant transverse polarization component, $f_\perp$, from the penguin annihilation diagrams. The $f_L$ for this decay is in agreement with the QCDF prediction~\cite{qcdfbtovv} but is significantly less than the previous PQCD prediction ($87\%$). From Table. \ref{tb:fl}, one also sees that for $B^0\to K^{*0}\omega$, our estimate for the longitudinal polarization fraction is in excellent agreement with the experimental data. We note that, for $B^0 \to \omega \omega$, our predicted longitudinal fraction is $66\%$, while the QCDF approach yields $94\%$~\cite{qcdfbtovv}, where the longitudinal contributions  highly dominate the amplitude. In QCDF, as in $B^0\to \rho^0\omega$, the penguin annihilation contributions are also tiny in $B^0 \to \omega\omega$. In PQCD, together with the hard scattering contributions, the considerable penguin annihilation contributions yield a different result. For the $B^0\to \phi\phi$, the previous PQCD prediction of the longitudinal polarization fraction is $65\%$~\cite{btovv}, while our updated longitudinal polarization fraction is given by $f_L\sim 1$, which is confirmed also in~\cite{YL}. For the $B^+\to K^{*0}\rho^+$ mode, our result is larger than the data, while in the QCDF framework, the central value is the same as the data, as this mode is used to extract the input parameters~\cite{qcdfbtovv}.

In this paragraph we shall discuss direct $CP$-asymmetries in the decays $B \to VV$ shown in Table \ref{tb:cp} and their current measurements. Though none of the current experimental measurements for the $CP$ asymmetries shown in Table \ref{tb:cp} is conclusive, they are in accord with our theoretical calculations. This, in turn, implies that the dominant strong phases in these channel estimated in our approach are in the right ball-park. From Table. \ref{tb:cp}, one also notes that the $CP$ asymmetries are large for the penguin dominant decays, but they are small for the color allowed tree-dominant decays and almost pure penguin-dominant processes, such as $K^{*0}\rho^+$ and $K^*\phi$. For $B^0\to \rho^0\omega$ decays, our prediction is about 60$\%$, while that of QCDF is only $3\%$. In PQCD, since the emission diagrams nearly cancel each other, the annihilation diagrams provide the dominant contributions. As direct $CP$ asymmetry is proportional to the interference between the tree and penguin contributions, the sizable interference makes the $CP$ asymmetry parameter large, reaching  $60\%$. For the $B^0\to \rho^0\rho^0/\omega\omega$ modes, the large penguin contributions from the chirally enhanced annihilation diagrams, which are at the same level as the tree contributions from the emission diagrams, make the the $CP$ asymmetry parameter as large as $70\%$. On the other hand, for pure annihilation type decay $B^0 \to K^{*0}\bar{K}^{*0}$, since there are no contributions from tree operators, it is natural to expect that the direct $CP$ asymmetry is practically zero. In summary, the entries in Tables~\ref{tb:br}, \ref{tb:fl} and \ref{tb:cp} show that for these $B \to VV$ decays our updated predictions are in good agreement with experiment, and, broadly speaking, are also in agreement with the QCDF predictions~\cite{qcdfbtovv}.

In Table \ref{tb:ft}, we give the predictions for the perpendicular polarization fraction, $f_\perp$, the relative phases,  $\phi_{\parallel}({\rm rad})$, $\phi_{\perp}({\rm rad})$, $\Delta\phi_{\parallel}(10^{-2}~{\rm rad})$, $\Delta\phi_{\perp}(10^{-2}~{\rm rad})$, and the $CP$ asymmetry parameters $A^{0}_{CP}$ and $A^{\perp}_{CP}$ for the $B \to VV$ decays for the first time in the PQCD framework. These remain to be confronted with the data.
In fact, these variables are already  experimentally measured in five channels:  $B^0(B^+) \to K^* (K^{*+}) \phi$ and $B_s \to \overline K^{*0}\phi$, $B_s \to \overline K^{*0}K^{*0}$ and $B_s \to \phi \phi$, which are shown in Table~\ref{tb:zui}. Our results are in good agreement with the data.

We now discuss the results for the $B_s\to VV$ decays. Since the initial and the final state distribution amplitudes (DAs) are the most important input parameters in the PQCD approach, our predictions for $B_s\to VV$ decays in Tables~\ref{tb:brbs}, \ref{tb:flbs}, \ref{tb:cpbs}, \ref{tb:ftbs} are almost the same as the predictions in~\cite{bsvv}, as the DAs we adopted here are similar to those used in~\cite{bsvv}, except for the DAs of the $\phi$ meson. For the $B_s\to \phi\phi$ decay, the central values of the branching ratio and the longitudinal polarization fraction estimated in~\cite{bsvv} are $35.3\times 10^{-6}$ and 61.9$\%$, respectively. It is apparent that neither the branching ratio nor the polarization fraction are in conformity with the experimental data, posted as $(19\pm5)\times 10^{-6}$ and $(34.8\pm4.6)\%$, respectively. With the updated DAs of $\phi$ meson, the current predictions of all the observables listed in Table~\ref{tb:zui} agree better with the data. This can be confirmed by the similar updates in $B_s\to \pi^+\pi^-$ and $B^0\to K^+K^-$ decays\cite{xiaowang}. Also, due to the terms proportional to the ratio $r_i^2=m_{\phi}^2/M_{B_s}^2$ in the denominators, which we keep, their influence is expected to be more pronounced, as $m_{\phi}$ is larger than the other light vector-meson masses. For the rest of the decay modes, the numerical values of the polarization fraction are basically consistent with the former PQCD predictions ~\cite{bsvv}.

From Tables~\ref{tb:brbs} and~\ref{tb:flbs}, we note that for the color-suppressed decays $B_s\to \overline{K}^{*0}\rho(\omega)$, the branching ratios in PQCD are smaller than in QCDF by a factor of 3 due to the near cancellation of the hard scattering contributions. On the other hand, chirally enhanced annihilation and the hard scattering diagrams enhance the transverse polarization contribution, making it comparable to  the longitudinal polarization one. For $B_s\to \omega\phi$, the pure emission mode, the $(S-P)(S+P)$ densities in the hard scattering diagrams also contribute a sizable transverse polarization component. For $B_s\to K^{*+}K^{*-}$, due to the large transverse polarization contribution from chirally enhanced annihilation diagrams, the longitudinal polarization fraction is as small as $40\%$, which is similar to $B_s\to K^{*0}\overline{K}^{*0}$. We also emphasize the measurements of the modes $B_s\to \overline{K}^{*0}\rho(\omega)$ and $B_s\to \omega \phi$ to distinguish among the competing dynamical models in the interpretation of the polarization anomaly.

Direct $CP$ asymmetries of $B_s\to VV$ decays are listed in Table~\ref{tb:cpbs}. We note that they are small for the penguin-dominant processes, since the interference between tree and penguin contributions due to the former are too small, which is opposite to the tree-dominant process $B_s\to K^{*-}\rho^+$, which also has a small direct $CP$ asymmetry. For $B_s \to \rho^{0}\phi$, QCDF predicts about $83\%$ for direct $CP$ asymmetry with large charming penguin contributions. But in the framework of PQCD, it is only $-4.3\%$ because this mode belongs to the pure emission-type processes. Hence, measurement of direct CP asymmetry in this mode will help us to distinguish the PQCD and the QCDF approaches.

As is well known, SU(3) symmetry relates a number of $B_s\to VV$ and $B_{u,d}\to VV$ processes, such as $B_s\to K^{*-}\rho^+$ and $B^0\to \rho^+\rho^- $. In the PQCD approach, presented here, this relation is well satisfied:
\begin{eqnarray}
&&\mathcal{B}(B_s\to K^{*-}\rho^+)=(24.0_{-8.7-1.4-2.4}^{+10.9+1.2+0.0})\sim \mathcal{B}(B^0\to \rho^+\rho^-)=(26.0_{-8.1-1.4-1.2}^{+10.1+1.4+1.5}),
\end{eqnarray}
in units of $10^{-6}$. On the other hand, SU(3)-breaking in the decay rates for $B\to K^*\phi$ and $B_s\to \phi\phi$ is significant, as can be seen in Table~\ref{tb:zui}. In the PQCD approach, the SU(3)-breaking effects are caused by the differences between the initial and final state wave functions, such as the shape parameter $\omega_{B}$ and $\omega_{B_s}$, as well as the decay constants of the $B$ and $B_s$ mesons, along with the Gegenbauer moments and the decay constants of the final vector mesons. They conspire to yield a cumulative 60\% SU(3)-breaking effect. Other SU(3)-breaking effects lie in between these two cases, as can be numerically calculated from the entries in various tables presented here.

$U$-spin symmetry, relating a number of $B_{(s)} \to h_1 h_2$ ($h_i$ are light mesons) has been advocated in the
literature~ \cite{rf}. For  $B_{(s)} \to VV$ decays, it has been  studied in~\cite{qcdfbsvv} and checked against
the explicit QCDF estimates, and  seems to hold well.  Since we have calculated the $B$ and $B_s$ decays to $VV$ in this work in the PQCD approach, we also check the $U$-spin symmetry in some representative decays studied in \cite{qcdfbsvv}:
\begin{eqnarray}
&&A_{CP}(B_s\to K^{*-}\rho^+)=-A_{CP}(B^0\to K^{*+}\rho^-)\frac{\mathcal{B}(B^0\to K^{*+}\rho^-)}{\mathcal{B}(B_s\to K^{*-}\rho^+)}\frac{\tau(B_s)}{\tau(B)},\nonumber\\
&&A_{CP}(B_s\to \bar{K}^{*0}\rho^0)=-A_{CP}(B^0\to K^{*0}\rho^0)\frac{\mathcal{B}(B^0\to K^{*0}\rho^0)}{\mathcal{B}(B_s\to \bar{K}^{*0}\rho^0)}\frac{\tau(B_s)}{\tau(B)},\nonumber\\
&&A_{CP}(B_s\to \rho^+\rho^-)=-A_{CP}(B^0\to K^{*+}K^{*-})\frac{\mathcal{B}(B^0\to K^{*+}K^{*-})}{\mathcal{B}(B_s\to \rho^+\rho^-)}\frac{\tau(B_s)}{\tau(B)},\nonumber\\
&&A_{CP}(B_s\to K^{*+}K^{*-})=-A_{CP}(B^0\to \rho^+\rho^-)\frac{\mathcal{B}(B^0\to \rho^+\rho^-)}{\mathcal{B}(B_s\to K^{*+}K^{*-})}\frac{\tau(B_s)}{\tau(B)}.
\end{eqnarray}
Using these $U$-spin relations as well as the branching ratios, the lifetimes of $B$ and $B_s$ mesons and the direct $CP$ asymmetries in $B$ decays, we can get the relevant direct $CP$ asymmetries in $B_s$ decays. This can be then compared with the explicit calculations in the PQCD approach to check whether the $U$-spin symmetry works well or not. We show this comparison in Table~\ref{tb:u-spin}, where the entries in the last two columns have to be compared with each other. We find that, within the calculational errors, the $U$-spin symmetry works well in the direct CP asymmetries in the PQCD approach as well.
\begin{table*}[h]
\begin{center}
 \caption{The direct $CP$ asymmetries ($\%$) in $B_s \to VV$ decays via $U$-spin relation together with the direct PQCD prediction. The branching ratios of $B$ and $B_s$ decays are in units of $10^{-6}$.}
 \begin{tabular}{lcclccc}
 \hline\hline\vspace{0.2cm}
 modes & Br &$A_{CP}$($\%$)&modes&Br&$A_{CP}$($\%$)($U$)&$A_{CP}$(PQCD) \\
 \hline\vspace{0.2cm}
$B^0\to K^{*+}\rho^-$&8.4&$24.5_{-1.5-3.4-0.6}^{+1.2+2.9+0.0}$&$B_s\to K^{*-}\rho^+$&24.0&-8.4&-$9.1_{-1.5-1.2-0.3}^{+1.4+1.0+0.2}$\\\vspace{0.2cm}
$B^0\to K^{*0}\rho^0$&3.3&$-8.9_{-0.6-2.8-1.0}^{+0.6+2.8+1.}$&$B_s\to \bar{K}^{*0}\rho^0$&0.40&72.3&$62.7_{-5.9-16.0-7.9}^{+6.4+10.5+7.5}$\\\vspace{0.2cm}
$B^0\to K^{*+}K^{*-}$&0.21&$29.8_{-5.7-9.5-4.7}^{+2.0+6.4+4.6}$&$B_s\to \rho^+\rho^-$&1.5&-4.1&-$2.9_{-1.1-1.3-0.2}^{+0.7+1.5+0.2}$\\\vspace{0.2cm}
$B^0\to \rho^+\rho^-$&26.0&$0.83_{-0.59-0.31-0.00}^{+0.50+0.66+0.00}$&$B_s\to K^{*+}K^{*-}$&5.4&-3.9&$8.8_{-8.9-2.9-0.2}^{+2.5+0.5+0.0}$\\
 \hline\hline
\end{tabular}\label{tb:u-spin}
\end{center}
\end{table*}

\section{SUMMARY}\label{summary}
In this paper, we have reexamined the branching ratios, polarization fractions, relative phases, and direct $CP$ asymmetries in $B_q\to VV$ ($q=u,d,s$) decays in the PQCD approach. Compared to the previous PQCD calculations, the updated longitudinal and transverse decay constants as well as the  Gegenbauer moments in the vector mesons wave functions have been adopted, which allows us to reduce the parametric uncertainties in the branching ratios and other observables. What concerns the predictions of the polarization fractions and their relative phases, we have kept track of the terms proportional to the ratio $r_i^2=m_{V_i}^2/m_B^2 (i=2,3)$, which have been ignored in some earlier estimations. In addition, we have studied the decay modes $B \to \rho(\omega)\phi$ that have not been explored before. For the observables $f_{\perp}$, $\phi_{\parallel}$, $\phi_{\perp}$, $\Delta \phi_{\parallel}$, $\Delta\phi_{\perp}$, $A^{0}_{CP}$, and $A_{CP}^{\perp}$, we have provided the first PQCD predictions. So, this work updates and goes beyond what is already known in this approach.

Our numerical results are listed in the Tables in the preceding section. For the well-measured decay modes, such as $B\to K^*\phi$ and $B_s\to \phi\phi$, the updated PQCD predictions for all the experimental observables fare better than the previous predictions in this approach, improving comparison  with experiments. In addition, in many $B(B_s)\to VV$ decays, our results agree with the updated QCDF predictions~\cite{qcdfbtovv,qcdfbsvv}, as well as with the experimental data. Yet, in some other cases, our predictions and those in QCDF differ and we have discussed  some of these decays, such as $B^0\to \rho^0(K^{*0})\omega$ involving the annihilation contributions.

For the tree dominated $B\to \rho\rho$ processes, our results respect the isospin triangle relations, while the experimental data, taken on the face value, shows significant isospin-violation. Our estimated decay rate and the polarization fraction in $B^0 \to \rho^0\rho^0$ are in good agreement with the Belle measurement, but not so compared to the BABAR data.  This calls for a refined measurement of $B\to \rho\rho$ decays in the future.

From the entries in Tables~\ref{tb:fl} and~\ref{tb:flbs}, we note that our updated longitudinal polarization fractions are in good agreement with the data and the predictions in the QCDF approach~\cite{qcdfbtovv,qcdfbsvv} in some topologies. But for the color-suppressed decay modes, $B^0\to\rho^0\rho^0$, $B^0\to \rho^0\omega$, $B_s\to\overline K^{*0}\rho^0$ and $B_s\to \overline K^{*0}\omega$, the longitudinal contributions dominate the decay amplitudes in the QCDF approach, while in this work, the transverse polarization contributions are comparable to the longitudinal polarization contributions, and are even dominant in the amplitude for $B^0\to\rho^0\rho^0$. This provides the possibility of distinguishing between these two approaches.

Table \ref{tb:cp} and \ref{tb:cpbs} list predictions of the $CP$ asymmetry parameters,which agree with the experimental data, wherever available,  and, generally, also with the QCDF predictions \cite{qcdfbtovv,qcdfbsvv} in some  topologies. For the color-suppressed decays, $B^0\to \rho^0\rho^0$, $B^0\to \omega\omega$, $B_s\to\overline K^{*0}\rho^0$ and $B_s\to \overline K^{*0}\omega$, both PQCD and QCDF predict large direct $CP$ asymmetries. But for $B^0\to \rho^0\omega$, the central value of the QCDF prediction is only 3$\%$, while the prediction of this work is about 60$\%$ due to the large annihilation contributions. For $B^+\to K^{*+}\omega$ and $B_s\to \rho\phi$ decays, which are almost purely dominated by penguin contributions, we predict very small $CP$ asymmetries, but QCDF predicts them to be of orders 0.56 and 0.83 respectively due to the charming penguins, which needs to be confirmed by experiments.

Our predictions for many $B_s^0\to VV$ decays basically agree with the previous PQCD predictions~\cite{bsvv}. But for a few penguin dominant decay modes, for example, $B_s\to \phi\phi$, $B_s\to \overline {K}^{*0}\phi$ and $B_s\to K^{*0}\overline K ^{*0}$, the improvements are significant, especially in the polarization fractions.

\section*{Acknowledgment}
We are grateful to Yue-Long Shen for useful discussions. This research was supported in part by the National  Science Foundation of China under
the Grant Nos.~11447032, 11175151, 11235005, 11205072, 11375208, 11228512,
 the Natural Science Foundation of Shandong province (ZR2014AQ013) and the Program for New Century Excellent Talents in University (NCET) by Ministry of Education of P. R. China (Grant No. NCET-13-0991).


\begin{appendix}
\section{Related Hard Functions}\label{app:a}
In this appendix, we summarize the functions that appear in the analytic formulas in the Section \ref{jiexi}. The first two diagrams in Fig.\ref{fig:diagram} are factorizable emission diagrams, whose hard scales $t_{a(b)}$ can be determined by
\begin{eqnarray}
t_a=\max\{\sqrt{x_3(1-r_2^2)}M_B,\;1/b_1,\;1/b_3\},
\end{eqnarray}
\begin{eqnarray}
t_b=\max\{\sqrt{x_1(1-r_2^2)}M_B,\;1/b_1,\;1/b_3\}.
\end{eqnarray}
The function $h_{ef}$ consists of two parts: the jet function $S_{t}(x)$ and the propagator of virtual quarks and gluons.
\begin{eqnarray}
h_{ef}(x_1,x_3,b_1,b_3)&=&K_0(\beta b_1)\left[\theta(b_1-b_3)I_0(\alpha b_3)K_0(\alpha b_1)\right.\nonumber\\
&&\left.+\theta(b_3-b_1)I_0(\alpha b_1)K_0(\alpha b_3)\right]S_t(x_3),
\end{eqnarray}
with $\alpha=\sqrt{x_3}M_B$ and $\beta=\sqrt{x_1x_3}M_B$. The jet function in the factorization formulas can be given as\cite{66094010}:
\begin{eqnarray}
S_t(x)=\frac{2^{1+2c}\Gamma(3/2+c)}{\sqrt{\pi}\Gamma(1+c)}\left[x(1-x)\right]^c,
\end{eqnarray}
with $c=0.4$. In the nonfactorizable contributions, due to the small numerical effect, we drop the jet function in the nonfactorizable emission diagrams and nonfactorizable annihilation diagrams\cite{plb555}.

The evolution factors $E_{ef}(t_{a})$ and $E_{ef}(t_{b})$ in the matrix elements are given by
\begin{eqnarray}
E_{ef}(t)\,=\,\alpha_{s}(t)\exp[-S_{B}(t)-S_{3}(t)].
\end{eqnarray}
The Sudakov exponents are defined as
\begin{eqnarray}
S_{B}(t)\,=\,s\left(x_{1}\frac{M_{B}}{\sqrt{2}},b_{1}\right)\,+\,\frac{5}{3}\int_{1/b_{1}}^{t}\frac{d\bar{\mu}}{\bar{\mu}}\gamma_{q}(\alpha_{s}(\bar{\mu})),
\end{eqnarray}
\begin{eqnarray}
S_{i}(t)\,=\,s\left(x_{i}\frac{M_{B}}{\sqrt{2}},b_i\right)\,+\,s\left((1-x_{i})\frac{M_{B}}{\sqrt{2}},b_i\right)
\,+\,2\int_{1/b_i}^{t}\frac{d\bar{\mu}}{\bar{\mu}}\gamma_{q}(\alpha_{s}(\bar{\mu})),
\end{eqnarray}
where the $s(Q,b)$ can be found in the Appendix A in the Ref.\cite{pqcd1}. $x_{i}$ is the momentum fraction of ``quark" in vector meson, with $i=2,3$.

For the rest of diagrams, the related functions are summarized as
follows:
\begin{eqnarray}
t_{c}=&&\max\{\sqrt{(1-r_2^2)x_3x_1}\,M_{B},\sqrt{\mid[(x_2-1)(1-r_3^2)+x_1)][r_2^2+x_3(1-r_2^{2})]\mid}\,M_{B},\nonumber\\
&&1/b_{1},1/b_{2}\},
\end{eqnarray}
\begin{eqnarray}
t_{d}=&&\max\{\sqrt{(1-r_2^2)x_3x_1}\,M_{B},\sqrt{\mid[x_2(r_3^2-1)+x_1)]x_3(1-r_2^{2})\mid}\,M_{B},\nonumber\\
&&1/b_{1},1/b_{2}\}.
\end{eqnarray}
\begin{eqnarray}
E_{enf}(t)\,=\,\alpha_{s}(t)\exp[-S_{B}(t)-S_{2}(t)-S_{3}(t)]| \,_{b_{1}=b_{3}}.
\end{eqnarray}
\begin{eqnarray}
h_{enf}(\alpha,\beta_i,b_{1},b_{2})\,&=&\,\left[\theta(b_{2}-b_{1})I_{0}(\alpha b_1)K_{0}(\alpha b_2)+\theta(b_{1}-b_{2})I_{0}(\alpha b_2)K_{0}(\alpha b_1)\right]\nonumber\\
&&\times \left\{\begin{array}{ll}
\frac{i\pi}{2}H_{0}^{(1)}\left(\sqrt{|\beta^2_i|}M_Bb_{2}\right),& \;\;\beta_i^2<0;\\
K_{0}\left(\beta_i M_B b_{2}\right),&\;\;\beta_i^{2}>0,
\end{array}\right.
\end{eqnarray}
with $i=1,2$ and
\begin{eqnarray}
\alpha &=&\sqrt{(1-r_2^2)x_3x_1}M_B,\\
\beta_{1}^{2}&=&[(x_2-1)(1-r_3^2)+x_1)][r_2^2+x_3(1-r_2^{2})],\\
\beta_2^{2}&=&[x_2(r_3^2-1)+x_1)]x_3(1-r_2^{2}),
\end{eqnarray}
The hard functions and the scales for factorizable annihilation diagrams Fig.(e) and (f) are
\begin{eqnarray}
&&t_{e}\,=\,\max\{ \alpha_1 M_{B},\beta M_B,1/b_{2},1/b_{3}\},\nonumber\\
&&t_{f}\,=\,\max\{\alpha_2 M_{B},\beta M_B,1/b_{2},1/b_{3}\},\\
&&E_{af}(t)\,=\,\alpha_{s}(t)\cdot \exp[-S_{2}(t)-S_{3}(t)],
\end{eqnarray}
\begin{eqnarray}
h_{af}(\alpha_i,\beta,b_{2},b_{3})\,&=&\,(\frac{i\pi}{2})^{2}H_{0}^{(1)}\left(\beta M_{B}b_{2}\right)\left[\theta(b_{2}-b_{3})H_{0}^{(1)}\left(\alpha_i M_{B}b_{2}\right)J_{0}\left(\alpha_i M_{B}b_{3}\right)\right.\nonumber\\
&&\left.+\theta(b_{3}-b_{2})H_{0}^{(1)}\left(\alpha_i M_{B}b_{3}\right)J_{0}\left(\alpha_i M_{B}b_{2}\right)\right]\cdot S_{t}(x_{3}),
\end{eqnarray}
with
\begin{eqnarray}
\alpha_1&=&\sqrt{1-x_3(1-r_2^2)}\\
\alpha_2&=&\sqrt{(1-r_2^2)[r_3^2+x_2(1-r_3^2)]},\\
\beta&=&\sqrt{[(1-r_2^2)(1-x_3)][r_3^2+x_2(1-r_3^2)]}.
\end{eqnarray}

For the nonfactorizable annihilation diagrams, the scales and the hard functions are
\begin{eqnarray}
t_g&=&\max\{\alpha M_{B}, \sqrt{|\beta_1|}M_{B}, 1/b_1,1/b_2\},\\
t_h&=&\max\{\alpha M_{B},\sqrt{|\beta_2|} M_{B},1/b_1,1/b_2\},\\
E_{anf}(t)\,&=&\,\alpha_{s}(t)\cdot
\exp[-S_{B}(t)-S_{2}(t)-S_{3}(t)]\mid\,_{b_{2}=b_{3}},
\end{eqnarray}
\begin{eqnarray}
h_{anf}(\alpha,\beta_i,b_{1},b_{2})\,&=&\,\frac{i\pi}{2}\left[\theta(b_{1}-b_{2})H_{0}^{(1)}\left(\alpha M_{B}b_{1}\right)J_{0}\left(\alpha M_{B}b_{2}\right)\right.\nonumber\\
&&\left.+\theta(b_{2}-b_{1})H_{0}^{(1)}\left(\alpha M_{B}b_{2}\right)J_{0}\left(\alpha M_{B}b_{1}\right)\right]\nonumber\\
&&\times \left\{\begin{array}{ll}
\frac{i\pi}{2}H_{0}^{(1)}\left(\sqrt{|\beta_i|}M_{B}b_{1}\right),&
\beta_{i}<0,\\
K_{0}\left(\sqrt{\beta_{i}}M_{B}b_{1}\right),& \beta_{i}>0,
\end{array}\right.
\end{eqnarray}
with $i=1,2$.
\begin{eqnarray}
\alpha&=&\sqrt{(1-x_3)(1-r_2^2)[r_3^2+x_2(1-r_3^2)]},\\
\beta_{1}&=&1-[(1-r_3^2)(1-x_2)-x_1][r_2^2+x_3(1-r_2^2)],\\
\beta_{2}&=&(1-r_2^2)(1-x_3)[x_1-x_2(1-r_3^2)-r_3^2].
\end{eqnarray}
\end{appendix}


\begin{thebibliography}{99}

\bibitem{qcdf1}
M. Beneke, J. Rohrer and D. S. Yang, Nucl. Phys. \textbf{B774}, 64 (2007).
\bibitem{qcdf2}
M. Bartsch, G. Buchalla and C. Kraus, arXiv:0810.0249 (2008).
\bibitem{qcdf3}
H. Y. Cheng and K. C. Yang, Phys. Rev. D \textbf{78} (2008) 094001, Erratum-ibid. D \textbf{ 79} (2009) 039903.
\bibitem{qcdfbtovv}
H. Y. Cheng and C. K. Chua, Phys. Rev. D \textbf{80}, 114008 (2009).
\bibitem{qcdfbsvv}
H. Y. Cheng and C. K. Chua, Phys. Rev. D \textbf{80}, 114026 (2009).
\bibitem{yang}
X. Q. Li, G. R. Lu and Y. D. Yang, Phys. Rev. D \textbf{68}, 114015 (2003), Erratum-ibid. D \textbf{71} 019902 (2005).
\bibitem{btovv}
Y. Li and C. D. L\"{u}, Phys. Rev. D \textbf{73}, 014024 (2006); H. W. Huang, C. D. L\"{u}, T. Morii, Y. L. Shen, G. L. Song, and J. Zhu, Phys. Rev. D \textbf{73}, 014011 (2006);
J. Zhu, Y. L Shen, and C. D. L\"{u}, Phys. Rev. D \textbf{72}, 054015 (2005); C. D. L\"{u}, Y. L. Shen, J. Zhu, Eur. Phys. J. C \textbf{41}, 311-317 (2005).
\bibitem{bsvv}
A. Ali, G. Kramer, Y. Li, C. D. L\"{u}, Y. L. Shen, W. Wang, and Y. M. Wang, Phys. Rev. D \textbf{76}, 074018 (2007).
\bibitem{jpg32}
J. Zhu, Y. L. Shen and C. D. L\"{u}, J. Phys. G \textbf{32}, 101-110 (2006).
\bibitem{hfag}
Heavy Flavour Averaging Group (HFAG);
Y.~Amhis {\it et al.}, arxiv:1412.7515,  and online update at http://www.slac.stanford.edu/xorg/hfag.
\bibitem{nfa}
M. Wirbel, B. Stech and M. Bauer, Z. Phys. C \textbf{29}, 637 (1985); B. Stech and M. wirbel, Z. Phys. C \textbf{34}, 103 (1987).
\bibitem{gaijin}
A. Ali and C. Greub, Phys. Rev. D \textbf{57}, 2996 (1998); G. Kramer, W. E. Palmer and H. Simma, Nucl. Phys. \textbf{B428}, 77 (1994); Z. Phys. C \textbf{66}, 429 (1995);
A. Ali, G. Kramer and C. D. L\"{u}, Phys. Rev. D \textbf{58}, 094009 (1998); Phys. Rev. D \textbf{59}, 014005 (1999); Y. H. Chen, H. Y. Cheng, B. Tseng and K. C. Yang, Phys. Rev. D \textbf{60} 094014 (1999).
\bibitem{qcdf}
M. Beneke, G. Buchalla, M. Neubert and C. T. Sachrajda, Phys. Rev. Lett. \textbf{83}, 1914 (1999); Nucl. Phys. \textbf{B591}, 313 (2000).
\bibitem{pdm}
M. Beneke and M. Neubert, Nucl. Phys. \textbf{B675}, 333 (2003).
\bibitem{pqcd}
Y. Y. Keum, H. N. Li and A. I. Sanda, Phys. Lett. B \textbf{504}, 6 (2001); C. D. L\"{u}, K. Ukai and M. Z. Yang, Phys. Rev. D \textbf{63}, 074009 (2001).
\bibitem{sect}
C. W. Bauer, D. Pirjol and I. W. Stewart, Phys. Rev. Lett. \textbf{87}, 201806 (2001).
\bibitem{lv23275}
C. D. L\"{u} and M. Z. Yang, Eur. Phys. J. C \textbf{23}, 275 (2002).
\bibitem{pqcd1}
Y.Y. Keum, H. N. Li, and A. I. Sanda, Phys. Lett. B \textbf{504}, 6 (2001); Phys. Rev. D \textbf{63}, 054008 (2001).
\bibitem{pqcd2}
C. D. L\"{u}, K. Ukai, and M. Z. Yang, Phys. Rev. D \textbf{63}, 074009.
\bibitem{pqcd3}
H. N. Li, Prog. Part. Phys. \textbf{51}, 85 (2003), and reference therein.
\bibitem{anni1}
C. D. L\"{u} and K. Ukai, Eur. Phys. J. C \textbf{28}, 305 (2003).
\bibitem{anni2}
Y. Li and C. D. L\"{u}, J. Phys. G \textbf{29}, 2115 (2003); High Energy Phys. Nucl. Phys.\textbf{27}, 1062 (2003).
\bibitem{pipi}
Y. Li, C. D. L\"{u}, Z. J. Xiao, and X. Q. Yu, Phys. Rev. D \textbf{70}, 034009 (2004).
\bibitem{dk}
R. H. Li, C. D. L\"{u}, and H. Zou, Phys. Rev. D \textbf{78}, 014018 (2008).
\bibitem{anniexp}
M. J. Morello \textit{et} \textit{al}. (CDF Collaboration), CDF public note, Report No. 10498,2011;
K. Nakamura \textit{et} \textit{al}. (Particle Data Group), J. Phys. G \textbf{37}, 075021 (2010).
\bibitem{rmp681125}
G. Buchalla, A. J. Buras and M. E. Lautenbacher, Rev. Mod. Phys. \textbf{68}, 1125 (1996).
\bibitem{prd58094009}
A. Ali, G.Kramer and C. D. L\"{u} in Ref.~\cite{gaijin}.

\bibitem{prd55and56}
C. H. Chang and H. N. Li, Phys. Rev. D \textbf{55}, 5577 (1997); T. W. Teh and H. N. Li, Phys. Rev. D \textbf{56}, 1615 (1997).
\bibitem{prd66094010}
H. N. Li, Phys. Rev. D \textbf{66}, 094010 (2002).
\bibitem{prd57443}
H. N. Li and B. Tseng, Phys. Rev. D \textbf{57}, 443 (1998)
\bibitem{lvepjc23275}
C. D. L\"{u} and M. Z. Yang, Eur. Phys. J. C \textbf{23}, 275-287
(2002).
\bibitem{bwave1}
A. G. Grozin and M. Neubert, Phys. Rev. D \textbf{55}, 272 (1997); M. Beneke and T. Feldmann, Nucl. Phys. \textbf{B592}, 3 (2001).
\bibitem{bwave2}
H. Kawamura, J. Kodaira, C. F. Qiao, and K. Tanaka, Phys. Lett. B \textbf{523}, 111 (2001); \textbf{536}, 344(E) (2002); Mod. Phys. Lett. A \textbf{18}, 799 (2003).
\bibitem{omg}
Y. Y. Keum, H. N. Li, Phys. Rev. D \textbf{63}, 074006 (2001); C. D. L\"{u}, M. Z. Yang, Eur. Phys. J. C \textbf{23}, 275 (2002); H. N. Li and H. L. Yu, Phys. Rev. D \textbf{53}, 2480 (1996).

\bibitem{bksphi}
H. N. Li, Phys. Lett. B \textbf{622}, 63 (2005).
\bibitem{vwave}
P. Ball, V. M. Braun, Y. Koike, and K. Tannka, Nucl. Phys. \textbf{B529}, 323 (1998); P. Ball and V. M. Braun, Nucl. Phys. \textbf{B543}, 201 (1999); P. Ball and R. Zwicky, Phys. Rev. D \textbf{71}, 014029 (2005).
\bibitem{jhep03}
P. Ball and G. W. Jones, J. High Energy Phys. 03 (2007) 069.
\bibitem{zhourui}
Zhou Rui, Gao Xiang-Dong, C. D. L\"{u}, Eur. Phys. J. C  \textbf{72}, 1923 (2012); Hsiang-nan Li, Yue-Long Shen, and Yu-Ming Wang, Phys. Rev. D \textbf{85}, 074004 (2012).

\bibitem{cptr}
B. H. Hong and C. D. L\"{u}, Sci. China G\textbf{49}, 357 (2006); H. W. Huang \textit{et al}.,Phys. Rev. D \textbf{73}, 014011 (2006); H. N. Li and S. Mishima, Phys. Rev. D \textbf{71}, 054025 (2005).
\bibitem{pdg}
J. Beringer \textit{et al}.,(Particle Data Group), Phys. Rev. D \textbf{86}, 010001 (2012).
\bibitem{bsksphi}
R. Aaij  \textit{et al}. (LHCb Collaboration), JHEP11(2013)092.
\bibitem{flpenguin}
A. L. Kaan, Phys. Lett. B \textbf{601}, 151 (2004).
\bibitem{flfsi}
H. Y. Cheng, C. K. Chua, and A. Soni, Phys. Rev. D \textbf{71}, 014030 (2005); P. Colangelo, F. De Fazio, and T. N. Pham, Phys. Lett. B \textbf{597}, 291 (2004).
\bibitem{flfft}
H. N. Li, Phys. Lett. B \textbf{622}, 63 (2005).
\bibitem{flnp}
C. S. Kim and Y. D. Yang, arXiv:hep-ph/0412364; S. Baek, A. Datta, P. Hamel, O. F. Hernandez, and D. London, Phys. Rev. D \textbf{72}, 094008 (2005); Q. Chang, X. Q. Li, and Y. D. Yang, J. High Energy Phys. 06 (2007) 038;
\bibitem{babar:rhorho}
B. Aubert \textit{et al}. (BABAR Collaboration), Phys. Rev. D \textbf{78}, 071104(R) (2008).
\bibitem{belle:rhorho}
P. Vanhoefer \textit{et al}. (Belle Collaboration), arXiv:1212.4015 [hep-ex].
\bibitem{belle:rhorho}
P. Vanhoefer \textit{et al}. (Belle Collaboration), arXiv:1212.4015 [hep-ex].
\bibitem{flpqcd}
H. N. Li and S. Mishima, Phys. Rev. D \textbf{71}, 054025 (2005).
\bibitem{cpks+rho-}
J. P. Lees \textit{et al}. (BABAR Collaboration) ,Phys. Rev. D \textbf{85}, 072005 (2012).
\bibitem{xiaowang}
Zhen-Jun Xiao, Wen-Fei Wang, Y.-Y. Fan, Phys. Rev. D \textbf{85},094003 (2012).
\bibitem{fllaw}
A. L. Kagan, Phys. Lett. B \textbf{601}, 151 (2004).
\bibitem{babar}
B. Aubert \textit{et al}. (BABAR Collaboration), Phys. Rev. D \textbf{78}, 071104 (2008).
\bibitem{belle}
C. C. Chiang \textit{et al}. (Belle Collaboration), Phys. Rev. D \textbf{78}, 111102 (2008).
\bibitem{Li:2006cva}
H.~n.~Li and S.~Mishima, Phys.\ Rev.\ D {\bf 73}, 114014 (2006)  [hep-ph/0602214];
H.~n.~Li, Private communications.
\bibitem{YL}
Y. Li, Phys. Rev. D \textbf{89}, 014003 (2014).
\bibitem{rf}
R. Fleisher, R. Knegjens, Eur. Phys. J. C \textbf{71}, 1789 (2011); R. Fleischer, Eur. Phys. J. C \textbf{10}, 299 (1999); K. D. Bruyn, R. Fleischer, arXiv:1412.6834 [hep-ph]; K. D. Bruyn, R. Fleischer, R. Knegjens, M. Merk, Nucl. Phys. \textbf{B868}, 351-367 (2013).
\bibitem{66094010}
H.-n. Li, Phys. Rev. D \textbf{66}, 094010 (2002).
\bibitem{plb555}
H.-n. Li and K. Ukai, Phys. Lett. B \textbf{555}, 197 (2003).
\end{thebibliography}
\end{document}